\documentclass[reqno,10pt,dvips]{amsart}

\usepackage{amssymb,mathptmx,cite,psfrag,eucal,array,setspace,geometry,amscd}
\usepackage[dvips]{graphicx}

\numberwithin{equation}{section}

\newcolumntype{C}{>{$}c<{$}} 
\allowdisplaybreaks

\newcommand{\alg}[1]{\mathfrak{#1}}
\newcommand{\group}[1]{\mathsf{#1}}

\newcommand{\func}[2]{#1 \left( #2 \right)}
\newcommand{\tfunc}[2]{#1 \bigl( #2 \bigr)}

\newcommand{\brac}[1]{\left( #1 \right)}
\newcommand{\sqbrac}[1]{\left[ #1 \right]}
\newcommand{\set}[1]{\left\{ #1 \right\}}
\newcommand{\st}{\mspace{5mu} : \mspace{5mu}}

\newcommand{\abs}[1]{\left| #1 \right|}

\newcommand{\ZZ}{\mathbb{Z}}
\newcommand{\NN}{\mathbb{N}}
\newcommand{\RR}{\mathbb{R}}

\newcommand{\ii}{\mathfrak{i}}

\newcommand{\killing}[2]{\kappa \bigl( #1 , #2 \bigr)}

\newcommand{\affine}[1]{\widehat{#1}}

\newcommand{\comm}[2]{\bigl[ #1 , #2 \bigr]}

\newcommand{\ket}[1]{\bigl\lvert #1 \bigr\rangle}
\newcommand{\braket}[2]{\bigl\langle #1 \bigr\rvert \bigl. #2 \bigr\rangle}
\newcommand{\bracket}[3]{\bigl\langle #1 \bigr\rvert #2 \bigl\lvert #3 \bigr\rangle} 

\newcommand{\normord}[1]{{} : #1 : {}} 

\newcommand{\IrrMod}[1]{\mathcal{L}_{#1}}

\newcommand{\AffIrrMod}[1]{\affine{\mathcal{L}}_{#1}}

\newcommand{\AffIndMod}[1]{\affine{\mathcal{W}}_{#1}}
\newcommand{\AffOthMod}[1]{\affine{\mathcal{E}}_{#1}}

\newcommand{\SLA}[2]{\alg{#1} \left( #2 \right)}
\newcommand{\AKMA}[2]{\affine{\alg{#1}} \left( #2 \right)}

\newcommand{\ch}[2]{\func{\chi_{#1}}{#2}}
\newcommand{\nch}[2]{\func{\widetilde{\chi}_{#1}}{#2}}

\newcommand{\fuse}{\mathbin{\times_{\! f}}}

\newcommand{\Jth}[2]{\vartheta_{#1} \bigl( #2 \bigr)}

\newcommand{\qnum}[2]{\left( #1 \right)_{#2}}

\newcommand{\eqnref}[1]{Equation~\eqref{#1}}
\newcommand{\eqnDref}[2]{Equations~\eqref{#1} and \eqref{#2}}

\newcommand{\secref}[1]{Section~\ref{#1}}

\newcommand{\appref}[1]{Appendix~\ref{#1}}
\newcommand{\figref}[1]{Figure~\ref{#1}}
\newcommand{\figDref}[2]{Figures~\ref{#1} and \ref{#2}}
\newcommand{\tabref}[1]{Table~\ref{#1}}

\newcommand{\cft}{conformal field theory}
\newcommand{\cfts}{conformal field theories}

\newcommand{\lcft}{logarithmic conformal field theory}
\newcommand{\lcfts}{logarithmic conformal field theories}
\newcommand{\WZW}{Wess-Zumino-Witten}
\newcommand{\ope}{operator product expansion}
\newcommand{\opes}{operator product expansions}
\newcommand{\hws}{highest weight state}
\newcommand{\hwss}{highest weight states}
\newcommand{\hwm}{highest weight module}
\newcommand{\hwms}{highest weight modules}

\DeclareMathOperator{\tr}{tr}

\begin{document}

\title{$\AKMA{sl}{2}_{-1/2}$ and the Triplet Model}

\author[D Ridout]{David Ridout}

\address[David Ridout]{
Centre de Recherches Math\'{e}matiques \\
Universit\'{e} de Montr\'{e}al \\
Qu\'{e}bec, Canada, H3C 3J7.
}

\email{ridout@crm.umontreal.ca}

\thanks{\today}

\begin{abstract}
Conformal field theories with $\AKMA{sl}{2}_{-1/2}$ symmetry are studied with a view to investigating logarithmic structures.  Applying the parafermionic coset construction to the non-logarithmic theory, a part of the structure of the triplet model is uncovered.  In particular, the coset theory is shown to admit the triplet $W$-algebra as a chiral algebra.  This motivates the introduction of an augmented $\AKMA{sl}{2}_{-1/2}$-theory for which the corresponding coset theory is precisely the triplet model.  This augmentation is envisaged to lead to a precise characterisation of the ``logarithmic lift'' of the non-logarithmic $\AKMA{sl}{2}_{-1/2}$-theory that has been proposed by Lesage \emph{et al}.
\end{abstract}

\maketitle

\onehalfspacing

\section{Introduction} \label{secIntro}

We continue here our investigation, begun in \cite{RidSL208}, of the fractional level \WZW{} model with chiral symmetry algebra $\AKMA{sl}{2}$ and level $k=-\tfrac{1}{2}$.  Such fractional level theories were originally proposed \cite{KenInf86} and studied \cite{KohFus88,LuMod89,BerFoc90,MatFra90,AwaFus92,RamNew93,FeiFus94,AndOpe95,PetFus96,FurAdm97,MatPri99} long ago as (supposedly) rational \cfts{}.  These works were inspired by attempts to generalise the coset construction of \cite{GodVir85} to non-unitary models and the discovery \cite{KacMod88b} that the characters of certain fractional level irreducible representations carry a finite-dimensional representation of the modular group.  Despite a significant amount of work however, fractional level theories remained poorly understood and their consistency as theories was regarded as questionable at best \cite{DiFCon97}.

More recently, a study of the $k=-\tfrac{4}{3}$ $\AKMA{sl}{2}$-theory revealed that the obstacle to understanding such theories was the assumption of rationality \cite{GabFus01} (in fact, non-rationality had already been pointed out in the vertex algebra literature \cite{AdaVer95}).  Using powerful algorithms to compute the abstract fusion of representations \cite{NahQua94,GabInd96}, it was shown that this fractional level theory is not rational; rather, it possesses an infinite spectrum of distinct representations.  Moreover, this theory was proven to be logarithmic, meaning that the Virasoro zero-mode $L_0$ failed to be diagonalisable.  Subsequent studies for $k=-\tfrac{1}{2}$ confirmed this lack of rationality and suggested that for certain levels one could have both a non-logarithmic and a logarithmic fractional level model \cite{LesSU202,LesLog04}.  These later works utilised the well known fact (see \cite{GurRel98} for example) that the $c=-1$ system of bosonic ghosts, which we shall hereafter refer to as the $\beta \gamma$ ghost system, exhibits an $\AKMA{sl}{2}_{-1/2}$ symmetry.

One of the stated aims of \cite{LesSU202} was to put the equivalence of the $\beta \gamma$ system and the $\AKMA{sl}{2}_{-1/2}$ model on a firm basis.  There, this problem was attacked with the help of an auxiliary free field realisation and some rather formidable computations of four-point correlators.  A motivation for the research reported in \cite{RidSL208} was to simplify and make precise this equivalence by working intrinsically, that is without needing any free fields.  This was achieved by using the elegant formalism of extended algebras developed in \cite{RidSU206,RidMin07} to realise the $\beta \gamma$ chiral algebra as the simple current extension of $\AKMA{sl}{2}_{-1/2}$.  This precise treatment allowed us to correct several errors in the literature and led to a complete description of the meaning of modular invariance in fractional level theories.

The rational (non-negative integer level) \WZW{} models have long been regarded as the fundamental building blocks of unitary (rational) \cft{}.  Their proposed fractional level counterpoints were intended to play a similar foundational r\^{o}le in constructing non-unitary (but still rational) \cfts{}.  What the results of \cite{GabFus01,LesSU202,LesLog04} suggest is that this r\^{o}le of fractional level \WZW{} models might be extended so as to also regard them as fundamental building blocks of (rational and quasi-rational) \lcft{}.  This is of not insignificant interest.  Logarithmic \cft{}, itself introduced almost twenty years ago \cite{RozQua92,GurLog93}, has recently undergone a resurgence of activity as various groups have started using it to describe so-called non-local observables in the continuum limit of statistical lattice models \cite{CarLog99,GurCon02,ReaExa01,GurCon04,PirPre04,PeaLog06,JenHei06,PeaSol07,ReaAss07,SimPer07,RasFus07,RidPer07,RasWEx08,RidPer08,PeaSol09,DubCon10} (see also \cite{FjeLog02,FeiMod06,EbeVir06,FeiLog06,GabLog06,GabFro08,AdaTri08,RidLog07,BusLus09,AdaLat09,GabFus09} for further mathematical and field-theoretic studies) and propose bridges with the theory of Schramm-Loewner evolution \cite{RidLog07,SaiGeo09}.  Moreover, it has recently been suggested that the \cfts{} dual to certain topological gravity theories on $\group{AdS}_3$ are logarithmic \cite{GruIns08}.

With this in mind, we are continuing our study of $\AKMA{sl}{2}_{-1/2}$-theories with the twin aims of understanding the logarithmic model proposed in \cite{LesLog04} and evaluating the (conjectured) r\^{o}les of these theories as building blocks from which we can construct other \lcfts{}.  To be sure, the closely related $\beta \gamma$ ghost system is commonly used as such a building block in free field realisations.  However, our philosophy is different.  Just as we have understood the relationship between $\AKMA{sl}{2}_{-1/2}$ and the $\beta \gamma$ ghosts by starting with the former rather than the latter (even though the latter is free), so we would like to achieve our stated aims without resorting to free field methods where possible.  As we will see, we are only partially successful in following this philosophy, but our deviations are not crippling to the overall idea.  The advantages of this approach are clarity and, more importantly, generality.  By developing methods suited to an intrinsic understanding of $\AKMA{sl}{2}_{-1/2}$, we expect to be able to apply these methods to other fractional level theories for which free field realisations are not obvious.

The article is organised as follows.  First, in \secref{secOld}, we summarise our notations and conventions and review some pertinent results on the non-logarithmic $\AKMA{sl}{2}_{-1/2}$-theory that were obtained in \cite{RidSL208}.  We take some time to explain how modular invariance is interpreted within this theory, emphasising in particular the r\^{o}le played by what we call the ``Grothendieck ring of modular characters'' and its relation to the fusion ring.  \secref{secCoset} commences with a detailed study of the $c=-2$ coset theory obtained from this $\AKMA{sl}{2}_{-1/2}$-theory and its obvious $\AKMA{u}{1}$-subtheory.  We identify the spectrum of the coset theory in terms of irreducible Virasoro modules and compute the multiplicities with which they appear.

We then note in \secref{secTriplet} that grouping the states of the coset theory appropriately gives rise to characters which match those of two of the irreducible modules of the celebrated \emph{triplet model} of \cite{GabRat96,GabLoc99}.  This is perhaps the best known, and certainly best understood, example of a \emph{rational} \lcft{}.  We subsequently verify that our coset theory admits the triplet algebra as a chiral algebra.  Here the computations become slightly unwieldy, so we depart from our usual philosophy and instead derive the chiral algebra of the corresponding coset of the $\beta \gamma$ system.  Naturally, this gives the simple current extension of the triplet algebra, the algebra of symplectic fermions \cite{KauCur95,KauSym00}.  The triplet algebra result follows straight-forwardly, leading to the relationships summarised in the following diagram:
\begin{equation*}
\begin{CD}
\text{Symplectic fermions} @<\text{$\AKMA{u}{1}$-coset}<< \text{$\beta \gamma$ ghosts} \\
@A\text{Simple current extension}AA @AA\text{Simple current extension}A \\
\text{Triplet model} @<\text{$\AKMA{u}{1}$-coset}<< \AKMA{sl}{2}_{-1/2}
\end{CD}
\mspace{10mu}.
\end{equation*}
The theories at left have $c=-2$ whereas those at right are $c=-1$.  Reversing the horizontal arrows in this diagram roughly corresponds to finding free field realisations.  However, we would like to stress once again that the methods we employ generalise much more readily.

Perhaps the most important realisation of this identification of the coset chiral algebra is that the coset spectrum is incomplete.  Indeed, the latter is not even modular invariant even though the parent theory is (in the sense described in \secref{secOld}).  We rectify this unsavoury feature in \secref{secAugment} by \emph{augmenting} the original $\AKMA{sl}{2}_{-1/2}$-theory with additional irreducible modules.  These turn out to be of the class known as relaxed \hwms{} \cite{FeiEqu98,SemEmb97}.  This motivates the introduction of a ``bigger'' $\AKMA{sl}{2}_{-1/2}$-theory, which we expect will be the sought-after intrinsic realisation of the ``logarithmic lift'' of \cite{LesLog04}.

However, we do not verify this here.  Rather, we prefer to defer the confirmation and investigation of the logarithmic nature of our augmentation to a companion article \cite{RidFus10}.  This is necessitated by the technical subtleties involved in determining the fusion rules of this new theory and fully analysing the mathematical structure of the indecomposable $\AKMA{sl}{2}$-modules so-obtained.  We instead conclude in \secref{secConc} with a brief comparison between what we have achieved and the constructions of \cite{LesLog04} and a proposal for how one can still accommodate the notion of modular invariance within our augmented theory.

\section{$\beta \gamma$ Ghosts and $\AKMA{sl}{2}_{-1/2}$} \label{secOld}

We first review for convenience certain features of the (non-logarithmic) fractional level $\AKMA{sl}{2}_{-1/2}$ model, as detailed in \cite{RidSL208}.  Our conventions for $\SLA{sl}{2}$ are given with respect to the basis $\set{e,h,f}$ for which the non-trivial commutation relations are
\begin{equation}
\comm{h}{e} = 2 e, \qquad \comm{e}{f} = -h \qquad \text{and} \qquad \comm{h}{f} = -2 f.
\end{equation}
The second of these relations perhaps deserves comment:  The basis we use is not the usual Cartan-Weyl one (in which $\comm{E}{F} = H$).  Rather, it is related to the Cartan-Weyl basis via $e = \tfrac{\ii}{2} \brac{E + F + \ii H}$, $h = \ii \brac{E-F}$ and $f = \tfrac{\ii}{2} \brac{E + F - \ii H}$.  In particular, the Killing form is determined by
\begin{equation} \label{eqnKilling}
\killing{h}{h} = 2 \qquad \text{and} \qquad \killing{e}{f} = -1,
\end{equation}
with all other combinations giving zero.  We make this choice of basis from the outset because it is tailored to the $\SLA{sl}{2 ; \RR}$ adjoint, $e^{\dag} = f$ and $h^{\dag} = h$, and it is with this adjoint that one derives the $\beta \gamma$ ghost system as an extended algebra.  Indeed, the usual $\SLA{su}{2}$ adjoint gives rise to a closely related, but mildly non-associative, extended symmetry algebra.

The above conventions for $\SLA{sl}{2}$ carry over to the affinisation $\AKMA{sl}{2}$ in the usual way.  With the central mode being replaced by the level $k = -\tfrac{1}{2}$, the non-trivial commutation relations are
\begin{equation}
\comm{h_m}{e_n} = 2 e_{m+n}, \quad \comm{h_m}{h_n} = -m \delta_{m+n,0}, \quad \comm{e_m}{f_n} = -h_{m+n} + \frac{1}{2} m \delta_{m+n,0} \quad \text{and} \quad \comm{h_m}{f_n} = -2 f_{m+n}.
\end{equation}
It follows from \eqnref{eqnKilling} that the energy-momentum tensor of the theory is given by
\begin{equation} \label{eqnDefT}
\func{T}{z} = \frac{1}{3} \brac{\frac{1}{2} \normord{\func{h}{z} \func{h}{z}} - \normord{\func{e}{z} \func{f}{z}} - \normord{\func{f}{z} \func{e}{z}}},
\end{equation}
and corresponds to central charge $c = -1$.  As one expects, the fields $\func{e}{z}$, $\func{h}{z}$ and $\func{f}{z}$ are affine primaries of conformal dimension $1$.

{
\psfrag{L0}[][]{$\AffIrrMod{0}$}
\psfrag{L1}[][]{$\AffIrrMod{1}$}
\psfrag{La}[][]{$\AffIrrMod{-1/2}$}
\psfrag{Lb}[][]{$\AffIrrMod{-3/2}$}
\psfrag{La*}[][]{$\AffIrrMod{-1/2}^*$}
\psfrag{Lb*}[][]{$\AffIrrMod{-3/2}^*$}
\psfrag{g}[][]{$\gamma$}
\psfrag{00}[][]{$\scriptstyle \brac{0,0}$}
\psfrag{aa}[][]{$\scriptstyle \brac{\tfrac{-1}{2},\tfrac{-1}{8}}$}
\psfrag{bb}[][]{$\scriptstyle \brac{\tfrac{1}{2},\tfrac{-1}{8}}$}
\psfrag{cc}[][]{$\scriptstyle \brac{-1,\tfrac{-1}{2}}$}
\psfrag{dd}[][]{$\scriptstyle \brac{1,\tfrac{-1}{2}}$}
\psfrag{ee}[][]{$\scriptstyle \brac{1,\tfrac{1}{2}}$}
\psfrag{ff}[][]{$\scriptstyle \brac{-1,\tfrac{1}{2}}$}
\psfrag{gg}[][]{$\scriptstyle \brac{\tfrac{-3}{2},\tfrac{-1}{8}}$}
\psfrag{hh}[][]{$\scriptstyle \brac{\tfrac{1}{2},\tfrac{7}{8}}$}
\psfrag{ii}[][]{$\scriptstyle \brac{\tfrac{3}{2},\tfrac{-1}{8}}$}
\psfrag{jj}[][]{$\scriptstyle \brac{\tfrac{-1}{2},\tfrac{7}{8}}$}
\psfrag{kk}[][]{$\scriptstyle \brac{-2,-1}$}
\psfrag{ll}[][]{$\scriptstyle \brac{0,1}$}
\psfrag{mm}[][]{$\scriptstyle \brac{2,-1}$}
\begin{figure}
\begin{center}
\includegraphics[width=15cm]{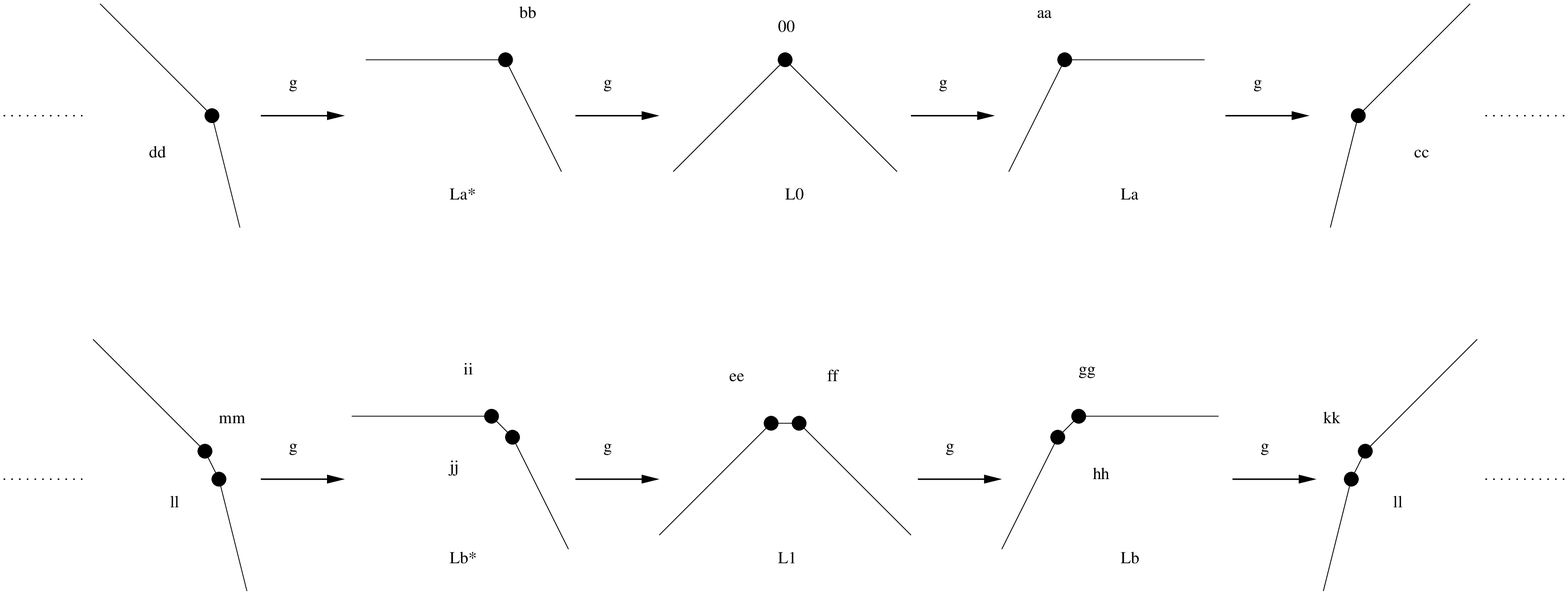}
\caption{Depictions of the modules appearing in the spectrum and the induced action of the spectral flow automorphism $\gamma$.  Each ``corner state'' is labelled by its $\func{\alg{sl}}{2}$-weight and conformal dimension (in that order).  Conformal dimensions increase from top to bottom and $\SLA{sl}{2}$-weights increase from right to left.} \label{figSpecFlow}
\end{center}
\end{figure}
}

The spectrum of this model consists of two infinite families of irreducible modules, $\tfunc{\gamma^{\ell}}{\AffIrrMod{0}}$ and $\tfunc{\gamma^{\ell}}{\AffIrrMod{1}}$, where $\ell \in \ZZ$.  Here, $\AffIrrMod{0}$ is the vacuum module of the theory, $\AffIrrMod{1}$ is a ``spin $\tfrac{1}{2}$'' module, and $\gamma$ denotes a \emph{spectral flow} automorphism of $\AKMA{sl}{2}$:
\begin{equation} \label{eqnSF}
\func{\gamma}{e_n} = e_{n-1}, \qquad \func{\gamma}{h_n} = h_n + \frac{1}{2} \delta_{n,0}, \qquad \func{\gamma}{f_n} = f_{n+1} \qquad \text{and} \qquad \func{\gamma}{L_0} = L_0 - \frac{1}{2} h_0 - \frac{1}{8}.
\end{equation}
As usual, such automorphisms allow us to twist any representation $\pi$ of $\AKMA{sl}{2}$ on a space (module) $\mathcal{M}$ by considering instead $\pi \circ \gamma^{-1}$ (the inverse power is conventional).  This is indeed a representation, but it will not be isomorphic to $\pi$ in general, even though the underlying representation space $\mathcal{M}$ has not changed.

We wish to concentrate on modules rather than representations in what follows.  It is therefore convenient to distinguish the above twisting at the level of modules by defining (somewhat artificially) the twisted module $\tfunc{\gamma^*}{\mathcal{M}}$.  As a vector space, this is identical to $\mathcal{M}$, but the action of $\AKMA{sl}{2}$ is different.  Specifically, if $\ket{v} \in \mathcal{M}$, so $\tfunc{\gamma^*}{\ket{v}} \in \tfunc{\gamma^*}{\mathcal{M}}$, then $\AKMA{sl}{2}$ is defined to act on the twisted module via
\begin{equation}
\tfunc{\pi^*}{J_n} \tfunc{\gamma^*}{\ket{v}} = \gamma^* \Bigl( \tfunc{\brac{\pi \circ \gamma^{-1}}}{J_n} \ket{v} \Bigr) \qquad \text{($J = e,h,f$).}
\end{equation}
Dropping the representations makes this more succinct:
\begin{equation}
J_n \tfunc{\gamma^*}{\ket{v}} = \func{\gamma^*}{\tfunc{\gamma^{-1}}{J_n} \ket{v}} \qquad \text{($J = e,h,f$).}
\end{equation}
We will moreover usually drop the superscript in $\gamma^*$ and instead speak of the automorphism $\gamma$ inducing twist maps, also denoted by $\gamma$, between modules.

We illustrate the families constituting the spectrum of the $\AKMA{sl}{2}_{-1/2}$-theory in \figref{figSpecFlow}.  Note that the conformal dimensions of the states of the modules with $\abs{\ell} > 1$ are not bounded below (in particular, they are not \hwms{}).  Practically, we will refer to the modules $\tfunc{\gamma^{\ell}}{\AffIrrMod{0}}$ and $\tfunc{\gamma^{\ell}}{\AffIrrMod{1}}$ with $\ell \neq 0$ as being twisted.  The untwisted modules $\AffIrrMod{0}$ and $\AffIrrMod{1}$ are then irreducible \hwms{} whose \hwss{} have respective $\SLA{sl}{2}$-weights $0$ and $1$, and respective conformal dimensions $0$ and $\tfrac{1}{2}$.  For later reference, we illustrate them with their first few weight space multiplicities in \figref{figGenMods}.

The only other \hwms{} in the spectrum are the (twisted) modules $\AffIrrMod{-1/2} = \tfunc{\gamma}{\AffIrrMod{0}}$ and $\AffIrrMod{-3/2} = \tfunc{\gamma}{\AffIrrMod{1}}$ which are generated by \hwss{} of $\SLA{sl}{2}$-weight $-\tfrac{1}{2}$ and $-\tfrac{3}{2}$, respectively, and conformal dimension $-\tfrac{1}{8}$.  Together with $\AffIrrMod{0}$ and $\AffIrrMod{1}$, these exhaust the so-called admissible modules of Kac and Wakimoto \cite{KacMod88b} (at $k=-\tfrac{1}{2}$).  However, $\AffIrrMod{0}$ and $\AffIrrMod{1}$ are self-conjugate representations, whereas $\AffIrrMod{-1/2}$ and $\AffIrrMod{-3/2}$ are not --- in general, the twisted module $\tfunc{\gamma^{\ell}}{\AffIrrMod{\lambda}}$ is conjugate to $\tfunc{\gamma^{-\ell}}{\AffIrrMod{\lambda}}$.  As usual, conjugation may be identified here with the standard (induced) action of the Weyl reflection $\mathsf{w}$ corresponding to the simple root of $\SLA{sl}{2}$:
\begin{equation} \label{eqnConjugation}
\func{\mathsf{w}}{e_n} = f_n, \qquad \func{\mathsf{w}}{h_n} = -h_n \qquad \text{and} \qquad \func{\mathsf{w}}{f_n} = e_n.
\end{equation}
On the other hand, the spectral flow \eqref{eqnSF} may be identified with a square root of an affine Weyl translation (and so $\gamma$ is an outer automorphism).  We mention that $\mathsf{w}$ and $\gamma$ generate the group of automorphisms of $\AKMA{sl}{2}$ which preserve the Cartan subalgebra.  They do not commute.

The fusion rules of our $\AKMA{sl}{2}$-modules are extremely simple, taking the form
\begin{equation} \label{eqnFR}
\tfunc{\gamma^{\ell_1}}{\AffIrrMod{\lambda}} \fuse \tfunc{\gamma^{\ell_2}}{\AffIrrMod{\mu}} = \tfunc{\gamma^{\ell_1 + \ell_2}}{\AffIrrMod{\lambda + \mu}}.
\end{equation}
Here, $\ell_1 , \ell_2 \in \ZZ$, whereas $\lambda$ and $\mu$ take value $0$ or $1$ and their sum is taken modulo $2$.  The computation of these fusion rules assumes that they respect the outer automorphisms of the chiral symmetry algebra, in this case the spectral flow, in the sense that 
\begin{equation} \label{eqnFusionAssumption}
\func{\gamma^{\ell_1}}{\mathcal{M}} \fuse \func{\gamma^{\ell_2}}{\mathcal{N}} = \func{\gamma^{\ell_1 + \ell_2}}{\mathcal{M} \fuse \mathcal{N}}
\end{equation}
for all (suitable) modules $\mathcal{M}$ and $\mathcal{N}$.  This assumption is very natural, but to our knowledge has never been proven, despite a significant amount of evidence in its favour.  We mention that \eqref{eqnFusionAssumption} does hold for the integrable modules of the rational \WZW{} models, though the standard proof is not at all elementary as it relies upon the Verlinde formula (see \cite[Sec.~16.1]{DiFCon97} for example).  We also note that \eqref{eqnFusionAssumption} does not hold if we replace $\gamma$ by the (inner) conjugation automorphism $\mathsf{w}$.  Instead, we have
\begin{equation} \label{eqnFusionConjugation}
\mathcal{M}^* \fuse \mathcal{N}^* = \func{\mathsf{w}}{\mathcal{M}} \fuse \func{\mathsf{w}}{\mathcal{N}} = \func{\mathsf{w}}{\mathcal{M} \fuse \mathcal{N}} = \brac{\mathcal{M} \fuse \mathcal{N}}^*,
\end{equation}
where we use ``$^*$'' to denote conjugate modules.

\begin{figure}
\begin{center}
\begin{tabular}{*{15}{@{}C@{\hspace{1mm}}}@{}C@{\hspace{7mm}}*{16}{@{}C@{\hspace{1mm}}}}
  &  &  &  &  &  &  & 1&  &  &  &  &  &  &  & &  &  &  &  &  &  &  & 1& 1&  &  &  &  &  &  &  \\
  &  &  &  &  &  & 1& 1& 1&  &  &  &  &  &  & &  &  &  &  &  &  & 1& 2& 2& 1&  &  &  &  &  &  \\
  &  &  &  &  & 1& 2& 3& 2& 1&  &  &  &  &  & &  &  &  &  &  & 1& 2& 4& 4& 2& 1&  &  &  &  &  \\
  &  &  &  & 1& 2& 5& 6& 5& 2& 1&  &  &  &  & &  &  &  &  & 1& 2& 5& 8& 8& 5& 2& 1&  &  &  &  \\
  &  &  & 1& 2& 5& 9&12& 9& 5& 2& 1&  &  &  & &  &  &  & 1& 2& 5&10&15&15&10& 5& 2& 1&  &  &  \\
  &  & 1& 2& 5&10&18&21&18&10& 5& 2& 1&  &  & &  &  & 1& 2& 5&10&19&27&27&19&10& 5& 2& 1&  &  \\
  & 1& 2& 5&10&20&31&38&31&20&10& 5& 2& 1&  & &  & 1& 2& 5&10&20&34&47&47&34&20&10& 5& 2& 1&  \\
 1& 2& 5&10&20&35&55&63&55&35&20&10& 5& 2& 1& & 1& 2& 5&10&20&36&60&79&79&60&36&20&10& 5& 2& 1 \\
\phantom{55}&\phantom{55}&\phantom{55}&\phantom{55}&\phantom{55}&\phantom{55}&\phantom{55}&\vdots&\phantom{55}&\phantom{55}&\phantom{55}&\phantom{55}&\phantom{55}&\phantom{55}&\phantom{55}& &\phantom{55}&\phantom{55}&\phantom{55}&\phantom{55}&\phantom{55}&\phantom{55}&\phantom{55}&\multicolumn{2}{C}{\vdots}&\phantom{55}&\phantom{55}&\phantom{55}&\phantom{55}&\phantom{55}&\phantom{55}&\phantom{55}\\
  &  &  &  &  &  &  &\AffIrrMod{0}&  &  &  &  &  &  &  & &  &  &  &  &  &  &  &\multicolumn{2}{C}{\AffIrrMod{1}}&  &  &  &  &  &  &  \\
\end{tabular}
\caption{The multiplicities of the weights of the generating representations of $\AKMA{sl}{2}_{-1/2}$ to grade $7$.  As in \figref{figSpecFlow}, the $\SLA{sl}{2}$-weight increases from right to left (the weights are even for $\AffIrrMod{0}$ and odd for $\AffIrrMod{1}$) and the conformal dimension increases from top to bottom (integral for $\AffIrrMod{0}$ and half-integral for $\AffIrrMod{1}$).} \label{figGenMods}
\end{center}
\end{figure}

Putting $\ell_1 = \ell_2 = 0$ and $\lambda = \mu = 1$ into \eqnref{eqnFR} gives $\AffIrrMod{1} \fuse \AffIrrMod{1} = \AffIrrMod{0}$, which means that $\AffIrrMod{1}$ is a simple current in the fusion ring.  Extending the chiral algebra by the corresponding simple current fields leads, and it is here that the choice of adjoint is key, to the $\beta \gamma$ ghost system.  This latter system is a \cft{} defined by two bosonic ghost fields of dimension $\tfrac{1}{2}$ whose \opes{} are
\begin{equation} \label{eqnGhostOPEs}
\func{\beta}{z} \func{\beta}{w} = 2 \func{e}{w} + \ldots, \quad \func{\beta}{z} \func{\gamma}{w} = \frac{-1}{z-w} + \func{h}{w} + \ldots \quad \text{and} \quad \func{\gamma}{z} \func{\gamma}{w} = 2 \func{f}{w} + \ldots
\end{equation}
The adjoint and non-trivial commutation relations of the modes are
\begin{equation} \label{eqnGhostAdjComm}
\beta_n^{\dag} = \gamma_{-n} \qquad \text{and} \qquad \comm{\gamma_m}{\beta_n} = \delta_{m+n,0}.
\end{equation}

The \opes{} \eqref{eqnGhostOPEs} make it clear that this ghost system is a free field theory.  It is therefore easy to obtain fermionic character formulae for its modules.  As each ghost module decomposes into two level $k = -\tfrac{1}{2}$ $\AKMA{sl}{2}$-modules, we deduce similar character formulae for the latter.  In particular, \cite[Eqn.~(9.18)]{RidSL208} yields
\begin{equation} \label{eqnChars}
\ch{\tfunc{\gamma^{\ell}}{\AffIrrMod{\lambda}}}{z ; q} = \tr_{\tfunc{\gamma^{\ell}}{\AffIrrMod{\lambda}}} z^{h_0} q^{L_0} = z^{-\ell / 2} q^{-\ell^2 / 8} \sum_{n \in \ZZ + \lambda / 2 \ } \sum_{m = \abs{n}}^{\infty} \frac{q^{m + \ell n}}{\qnum{q}{m-n} \qnum{q}{m+n}} z^{2n},
\end{equation}
where $\qnum{q}{n} = \prod_{j=1}^n \brac{1-q^j}$ as always.  This should be compared to the standard bosonic character formulae (reproduced in \cite[Eqns.~(6.4)--(6.5)]{RidSL208} for example):
\begin{equation} \label{eqnBosonicChars}
\ch{\tfunc{\gamma^{\ell}}{\AffIrrMod{\lambda}}}{z ; q} = q^{-1/24} \frac{\displaystyle \sum_{r \in \ZZ + \brac{3 \ell + 2 \lambda + 2} / 12} z^{6r} q^{6r^2} - \sum_{r \in \ZZ + \brac{3 \ell - 2 \lambda - 2} / 12} z^{6r} q^{6r^2}}{\displaystyle \sum_{r \in \ZZ + \brac{2 \ell + 1} / 4} z^{4r} q^{2r^2} - \sum_{r \in \ZZ + \brac{2 \ell - 1} / 4} z^{4r} q^{2r^2}}.
\end{equation}
Both the bosonic and fermionic forms converge for $\abs{q} < 1$ and must be expanded in the annulus
\begin{equation} \label{eqnConvergenceRegion}
\abs{q}^{-\ell + 1} < \abs{z}^2 < \abs{q}^{-\ell - 1}
\end{equation}
in the complex $z$-plane so that the correct weight multiplicities of the module are generated \cite{LesSU202}.

We emphasise however that the $\beta \gamma$ ghost theory is \emph{not} equivalent to the $\AKMA{sl}{2}_{-1/2}$-theory under consideration.  The former is obtained by formally extending the chiral algebra of the latter by the fields of the $\AKMA{sl}{2}$-module $\AffIrrMod{1}$.  But the modular invariants of the $\AKMA{sl}{2}_{-1/2}$-theory were completely classified in \cite[Sec.~10]{RidSL208} and it turns out that the fields of $\AffIrrMod{1}$ are \emph{never} coupled to the (antiholomorphic) identity field.  The ``equivalence'' of the theories can therefore only be regarded as pertaining to their chiral halves, not as full \cfts{}.  This is not insubstantial however.  For example, it allows us to identify certain \emph{chiral} correlation functions in the two theories (although we would then have to ``glue'' the holomorphic and antiholomorphic results differently when constructing the full correlators of the two inequivalent theories).

It remains to discuss the modular invariance of the theory.  The theory is only \emph{quasi-rational} (in the sense of Nahm \cite{NahQua94}) so its modular properties are significantly more subtle than those of rational \cfts{}.  We will therefore take some time to emphasise the differences.  First, we normalise the characters (in the standard way) by multiplying by $y^k q^{-c/24} = y^{-1/2} q^{1/24}$.  They may then be expressed in terms of Jacobi theta functions and the Dedekind eta function:
{
\setlength{\extrarowheight}{2mm}
\begin{equation} \label{eqnNormChs}
\nch{\tfunc{\gamma^{\ell}}{\AffIrrMod{\lambda}}}{y ; z ; q} = 
\begin{cases}
\displaystyle \frac{y^{-1/2} \func{\eta}{q}}{2} \sqbrac{\frac{\brac{-1}^{\ell / 2}}{\Jth{4}{z ; q}} + \frac{\brac{-1}^{\lambda}}{\Jth{3}{z ; q}}} & \text{if $\ell$ is even,} \\
\displaystyle \frac{y^{-1/2} \func{\eta}{q}}{2} \sqbrac{\frac{\brac{-1}^{\brac{\ell + 1} / 2} \ii}{\Jth{1}{z ; q}} + \frac{\brac{-1}^{\lambda}}{\Jth{2}{z ; q}}} & \text{if $\ell$ is odd.}
\end{cases}
\end{equation}
}%
The normalised characters therefore span a four-dimensional vector space, despite the fact that they are supposed to collectively encode the weight space multiplicities of an infinite number of modules.  Modules with the same character are only distinguished by their (disjoint) annuli of convergence \eqref{eqnConvergenceRegion} in the $z$-plane.  However, these annuli are not preserved under modular transformations, so we can only define an action of the modular group if we agree to ignore convergence regions.\footnote{Of course, we could just conclude from this observation that modular transformations are meaningless for the \cft{} under consideration.  We regard this conclusion as unsatisfactory because the proposal here leads to the well-known finite-dimensional representation of the modular group and a perfectly reasonable Verlinde formula.  However, the relevance of this proposal to physics remains to be determined.  In particular, it is not clear at present if such a modular invariant guarantees the consistency of the theory on a torus.}  It follows that the modular properties of the theory will be blind to such distinctions, so only four independent characters can be relevant to such considerations.  We will refer to the characters \eqref{eqnNormChs}, \emph{sans} convergence region, as \emph{modular characters} in what follows.

To summarise, we have a bijective correspondence between modules and characters with a given region of convergence.  To consider modular properties, we are now forced to ``forget'' these convergence regions, thereby losing the bijectivity.  What we gain is that the so-obtained modular characters are now, artificially, defined (almost) everywhere in the $z$-plane, so that they admit a well-defined action of the modular group.  Formally, we are defining a projection from the abelian group (with operation $\oplus$) generated by the $\tfunc{\gamma^{\ell}}{\AffIrrMod{\lambda}}$, the \emph{Grothendieck group of modules}, to the abelian group generated by the corresponding normalised characters \eqref{eqnNormChs}, the \emph{Grothendieck group of modular characters}.  One can easily check that the kernel of this projection is precisely generated by the modules of the form
\begin{equation} \label{eqnGenKer}
\tfunc{\gamma^{\ell \pm 1}}{\AffIrrMod{0}} \oplus \tfunc{\gamma^{\ell \mp 1}}{\AffIrrMod{1}}.
\end{equation}

Now, the Grothendieck group of modules comes equipped with a multiplication --- fusion.  We may therefore speak of the Grothendieck ring of modules, or in more standard terminology, the fusion ring.  But it is easy to check from \eqnref{eqnFR} that the kernel of the projection discussed above is an ideal of the fusion ring, so the fusion operation descends naturally to the Grothendieck group of modular characters (fusion is therefore well-defined at the level of these characters).  This latter group may thus be given a ring structure, so we will refer to it as the \emph{Grothendieck ring of modular characters}.

The above discussion makes it clear that the Grothendieck ring of modular characters is a quotient of the true fusion ring of our theory.  However, it does not possess some of the nice features that one would expect of a fusion ring.  In particular, the structure constants of a fusion ring (the fusion coefficients) are non-negative integers in the canonical basis, and the conjugation automorphism $\mathsf{w}$ acts as a permutation there.  Neither of these statements holds in the Grothendieck ring of modular characters (note that $\mathsf{w}$ has a well-defined action on this quotient).  They are spoiled by the consistent appearance of negative integer coefficients, which we can trace back to the ``$\oplus$'' in \eqnref{eqnGenKer}.

Nevertheless, it is the Grothendieck ring of modular characters on which the modular group acts, \emph{not} the fusion ring.  The $S$ and $T$-matrices of the theory are therefore $4$ by $4$ matrices, and turn out to be symmetric and unitary as one would hope.  But, as has been known for a long time, $S^2$ is not a permutation matrix and the Verlinde formula,
\begin{equation}
\mathsf{N}_{ij}^{\hphantom{ij} k} = \sum_{r} \frac{S_{ir} S_{jr} S_{kr}^*}{S_{0r}},
\end{equation}
gives positive and negative integer ``fusion'' coefficients.  However, one easily verifies that $S^2$ is precisely the matrix representing conjugation in the Grothendieck ring of modular characters and that the $\mathsf{N}_{ij}^{\hphantom{ij} k}$ are precisely the structure constants of this ring.  This realisation \cite{RidSL208} fully resolved the puzzle of ``negative fusion coefficients'', at least for $\AKMA{sl}{2}_{-1/2}$, and it is clear that the story should be analogous for other fractional level theories.

\section{The Coset Theory} \label{secCoset}

We will construct the coset theory
\begin{equation} \label{eqnCosetModel}
\frac{\AKMA{sl}{2}_{-1/2}}{\AKMA{u}{1}},
\end{equation}
where the $\AKMA{u}{1}$ subtheory is generated by the Virasoro primary field $\func{h}{z}$.  This is then a fractional level analogue of the parafermion theories of \cite{ZamNon85}.  The $\AKMA{u}{1}$ subtheory has energy-momentum tensor given by
\begin{equation} \label{eqnDeft}
\func{t}{z} = -\frac{1}{2} \normord{\func{h}{z} \func{h}{z}}
\end{equation}
and central charge $1$, so the coset theory is a $c = -2$ theory with energy-momentum tensor
\begin{equation} \label{eqnDefTT}
\func{\mathbb{T}}{z} = \func{T}{z} - \func{t}{z} = \frac{2}{3} \normord{\func{h}{z} \func{h}{z}} - \frac{1}{3} \normord{\func{e}{z} \func{f}{z}} - \frac{1}{3} \normord{\func{f}{z} \func{e}{z}}.
\end{equation}
It follows that the Virasoro zero-mode in the coset theory takes the form
\begin{equation} \label{eqnDefCosetL0}
\mathbb{L}_0 = L_0 + \frac{1}{2} h_0^2 + \sum_{r>0} h_{-r} h_r.
\end{equation}
The states of the coset theory may then be realised as the $\AKMA{u}{1}$-\hwss{} of the $\AKMA{sl}{2}_{-1/2}$-theory.  The coset construction \cite{GodVir85} guarantees that these states carry a representation of the Virasoro algebra $\alg{vir}$.

As $\AKMA{u}{1}$-Verma modules are always irreducible, it is easy to decompose the $\AKMA{sl}{2}$-modules into their $\AKMA{u}{1}$ constituents (this of course assumes that these constituents are actually \hwms{}).  The character of a $\AKMA{u}{1}$-Verma module of $\SLA{sl}{2}$-weight $2n \in \ZZ$ is
\begin{equation} \label{eqnU(1)Char}
\frac{q^{-2n^2}}{\qnum{q}{\infty}} = q^{-2n^2} \sqbrac{1 + q + 2 q^2 + 3 q^3 + 5 q^4 + 7 q^5 + 11 q^6 + 15 q^7 + \ldots},
\end{equation}
so by subtracting these multiplicities appropriately from those of \figref{figGenMods}, we arrive at a picture of the multiplicities of $\AKMA{u}{1}$-\hwss{} in the $\AKMA{sl}{2}$-modules $\AffIrrMod{0}$ and $\AffIrrMod{1}$.  This is illustrated in \figref{figu(1)Mults}.  These $\AKMA{u}{1}$-\hwss{} therefore carry a $c = -2$ representation of $\alg{vir}$ whose character we can now read off (at least to grade $7$).

\begin{figure}
\begin{center}
\begin{tabular}{*{15}{@{}C@{\hspace{1mm}}}@{}C@{\hspace{7mm}}*{16}{@{}C@{\hspace{1mm}}}}
  &  &  &  &  &  &  & 1&  &  &  &  &  &  &  & &  &  &  &  &  &  &  & 1& 1&  &  &  &  &  &  &  \\
  &  &  &  &  &  & 1& 0& 1&  &  &  &  &  &  & &  &  &  &  &  &  & 1& 1& 1& 1&  &  &  &  &  &  \\
  &  &  &  &  & 1& 1& 1& 1& 1&  &  &  &  &  & &  &  &  &  &  & 1& 1& 1& 1& 1& 1&  &  &  &  &  \\
  &  &  &  & 1& 1& 2& 2& 2& 1& 1&  &  &  &  & &  &  &  &  & 1& 1& 2& 2& 2& 2& 1& 1&  &  &  &  \\
  &  &  & 1& 1& 2& 2& 3& 2& 2& 1& 1&  &  &  & &  &  &  & 1& 1& 2& 3& 3& 3& 3& 2& 1& 1&  &  &  \\
  &  & 1& 1& 2& 3& 4& 4& 4& 3& 2& 1& 1&  &  & &  &  & 1& 1& 2& 3& 4& 5& 5& 4& 3& 2& 1& 1&  &  \\
  & 1& 1& 2& 3& 5& 5& 6& 5& 5& 3& 2& 1& 1&  & &  & 1& 1& 2& 3& 5& 6& 7& 7& 6& 5& 3& 2& 1& 1&  \\
 1& 1& 2& 3& 5& 6& 8& 8& 8& 6& 5& 3& 2& 1& 1& & 1& 1& 2& 3& 5& 7& 9&10&10& 9& 7& 5& 3& 2& 1& 1 \\
\phantom{55}&\phantom{55}&\phantom{55}&\phantom{55}&\phantom{55}&\phantom{55}&\phantom{55}&\vdots&\phantom{55}&\phantom{55}&\phantom{55}&\phantom{55}&\phantom{55}&\phantom{55}&\phantom{55}& &\phantom{55}&\phantom{55}&\phantom{55}&\phantom{55}&\phantom{55}&\phantom{55}&\phantom{55}&\multicolumn{2}{C}{\vdots}&\phantom{55}&\phantom{55}&\phantom{55}&\phantom{55}&\phantom{55}&\phantom{55}&\phantom{55}\\
  &  &  &  &  &  &  &\AffIrrMod{0}&  &  &  &  &  &  &  & &  &  &  &  &  &  &  &\multicolumn{2}{C}{\AffIrrMod{1}}&  &  &  &  &  &  &  \\
\end{tabular}
\caption{The multiplicities of the $\AKMA{u}{1}$-\hwss{} in the generating representations of $\AKMA{sl}{2}_{-1/2}$ to grade $7$.  The $\SLA{sl}{2}$-weight and conformal dimension are as in \figDref{figSpecFlow}{figGenMods} (the dimensions are with respect to $\AKMA{sl}{2}_{-1/2}$, not its coset).} \label{figu(1)Mults}
\end{center}
\end{figure}

For example, restricting to the subspace of $\AffIrrMod{0}$ whose states have vanishing $\SLA{sl}{2}$-weight, \figref{figGenMods} gives its character (to order $7$) as
\begin{equation}
1 + q + 3 q^2 + 6 q^3 + 12 q^4 + 21 q^5 + 38 q^6 + 63 q^7 + \ldots
\end{equation}
Since the highest state is obviously a $\AKMA{u}{1}$-\hws{} of dimension $0$, we can subtract the multiplicities of \eqnref{eqnU(1)Char} to get
\begin{equation}
q^2 + 3 q^3 + 7 q^4 + 14 q^5 + 27 q^6 + 48 q^7 + \ldots
\end{equation}
This indicates that there must be another $\AKMA{u}{1}$-\hws{} of dimension $2$.  Repeating this process, we obtain the multiplicities (for the zero-weight subspace) indicated in \figref{figu(1)Mults}:
\begin{equation} \label{eqnU(1)Mults}
1 + q^2 + 2 q^3 + 3 q^4 + 4 q^5 + 6 q^6 + 8 q^7 + \ldots
\end{equation}

Now the highest of these $\AKMA{u}{1}$-\hwss{} must be a $\alg{vir}$-\hws{} of dimension $0$ (under the coset Virasoro action).  There is no $\AKMA{u}{1}$-\hws{} with weight $0$ and dimension $1$, so this $\alg{vir}$-\hws{} can be identified as the vacuum of the coset theory.  Indeed, as $c = -2$, we can therefore conclude from the structure theory of highest weight Virasoro modules \cite{FeiSke82} that this vacuum generates an irreducible module with character
\begin{equation}
\frac{1-q}{\qnum{q}{\infty}} = 1 + q^2 + q^3 + 2 q^4 + 2 q^5 + 4 q^6 + 4 q^7 + \ldots
\end{equation}
Subtracting these multiplicities from those of \eqnref{eqnU(1)Mults}, we deduce that there must exist a $\alg{vir}$-\hws{} of dimension $3$.  It too is seen to generate an irreducible module (by checking that its singular descendant at grade $3$ vanishes), so we find that the $\SLA{sl}{2}$-weight zero states decompose into irreducible $\alg{vir}$-modules of dimension $0$ and $3$ (and probably others).

This decomposition procedure is tedious, but easily implemented on a computer.  We can therefore explore the coset theory in terms of its Virasoro modules to some depth.  Before describing the results of such investigation, let us just note the simple case in which we analyse the highest state in the subspace of states whose $\SLA{sl}{2}$-weights are constant, $2n \in \ZZ$ say.  Such a state is obviously a \hws{} in the coset theory and \eqnref{eqnDefCosetL0} indicates that its conformal dimension will be $\abs{n} - \brac{-\tfrac{1}{2} \, 4n^2} = \abs{n} \brac{2 \abs{n} + 1}$.

By repeating this decomposition analysis deeper in the $\AKMA{sl}{2}$-modules, we are led to a precise conjecture:  The set of $\AKMA{u}{1}$-\hwss{} with given $\SLA{sl}{2}$-weight $2n \in \ZZ$ decomposes as a $\alg{vir}$-module into the irreducibles
\begin{equation} \label{eqnSumModules}
\sideset{}{'} \bigoplus_{m = 2 \abs{n}}^{\infty} \IrrMod{m \brac{m+1} / 2}.
\end{equation}
Here, $\IrrMod{h}$ denotes the irreducible ($c = -2$) $\alg{vir}$-module whose \hws{} has conformal dimension $h$, and the prime indicates that the sum index $m$ increases by $2$.  We emphasise that $n$ may take half-integer values, so \eqnref{eqnSumModules} describes the coset decomposition of both $\AKMA{sl}{2}$-modules $\AffIrrMod{0}$ and $\AffIrrMod{1}$.  It now follows that the character of this infinite sum of modules is then\footnote{Strictly speaking, it is this character which we conjecture based on the above analysis.}
\begin{equation} \label{eqnCosetChar1}
\sideset{}{'} \sum_{m = 2 \abs{n}}^{\infty} \frac{q^{m \brac{m+1} / 2} - q^{\brac{m+1} \brac{m+2} / 2}}{\qnum{q}{\infty}} = \sum_{m = 2 \abs{n}}^{\infty} \brac{-1}^{m - 2n} \frac{q^{m \brac{m+1} / 2}}{\qnum{q}{\infty}}
\end{equation}
(the unprimed sum on the right hand side increases by $1$ as per usual).  Comparing with \eqnref{eqnChars}, we see that our conjecture will be proven if we can demonstrate the following equality:
\begin{equation} \label{eqnCombId1}
\sum_{m = \abs{n}}^{\infty} \frac{q^m}{\qnum{q}{m-n} \qnum{q}{m+n}} = \frac{q^{-2 n^2}}{\qnum{q}{\infty}^2} \sum_{m = 2 \abs{n}}^{\infty} \brac{-1}^{m - 2n} q^{m \brac{m+1} / 2} \qquad \text{($2n \in \ZZ$).}
\end{equation}
The additional factor of $q^{-2 n^2} / \qnum{q}{\infty}$ on the right hand side precisely accounts for the characters \eqref{eqnU(1)Char} of the $\AKMA{u}{1}$-modules with respect to which we have decomposed our theory.  We verify \eqnref{eqnCombId1} in \appref{appComb}.

We could also decompose the twisted modules $\tfunc{\gamma^{\ell}}{\AffIrrMod{\lambda}}$ with $\ell \neq 0$ into $\AKMA{u}{1}$-modules and determine the corresponding coset $\alg{vir}$-representations.  However, we would quickly discover that there is little point to this exercise, as the coset representations obtained are the same, regardless of the value of $\ell$.  This should not be surprising as the Virasoro algebra does not admit any non-trivial spectral flow automorphisms.  To verify this properly, note that under the action of the spectral flow automorphism $\gamma$ (given in \eqnref{eqnSF}), $\SLA{sl}{2}$-weights are only shifted by a constant, in fact by the level $k = -\tfrac{1}{2}$.  Spectral flow therefore maps a set of states of constant $\SLA{sl}{2}$-weight to another set of states of constant $\SLA{sl}{2}$-weight, preserving the multiplicities.  The decomposition into $\AKMA{u}{1}$-modules will therefore be identical.  It remains only to check the conformal dimensions of the coset (Virasoro) \hwss{}.  But, applying $\gamma$ to \eqnref{eqnDefCosetL0} gives
\begin{equation}
\tfunc{\gamma}{\mathbb{L}_0} = \func{\gamma}{L_0 + \frac{1}{2} h_0^2 + \sum_{r>0} h_{-r} h_r} = L_0 - \frac{1}{2} h_0 - \frac{1}{8} + \frac{1}{2} \brac{h_0 + \frac{1}{2}}^2 + \sum_{r>0} h_{-r} h_r = \mathbb{L}_0,
\end{equation}
so the coset dimensions do not change under spectral flow.

In summary then, we have derived the spectrum of the (chiral) coset theory.  With the usual field identifications derived from chiral algebra automorphisms (spectral flow), we can restrict our attention to the decomposition of the untwisted modules $\AffIrrMod{0}$ and $\AffIrrMod{1}$.  The spectrum then consists of the irreducible $\alg{vir}$-modules $\IrrMod{m \brac{m+1} / 2}$ for $m \in \NN$, and each such module appears with multiplicity $m+1$.  However, these $m+1$ copies are not completely identical.  Because the states of the coset theory are $\AKMA{u}{1}$-\hwss{} of $\AKMA{sl}{2}$-modules, they come equipped with an extra quantum number, the $\SLA{sl}{2}$-weight (or equivalently, the $\func{u}{1}$-charge).  It is therefore more honest to say that the $m+1$ copies of the $\alg{vir}$-module $\IrrMod{m \brac{m+1} / 2}$ are distinguished by their $\SLA{sl}{2}$-weights, which are $m$, $m-2$, $\ldots$, $-m+2$ and $-m$.

\section{The Triplet Model} \label{secTriplet}

Readers familiar with $c = -2$ \cfts{} will no doubt recognise the conformal dimensions $\tfrac{1}{2} m \brac{m+1}$ which appear in the spectrum of our coset theory.  Indeed, these constitute the first column of the extended Kac table for this central charge, a part of which is reproduced in \tabref{tabKacTable} (the Kac table proper is empty).  This lists the conformal dimensions of the \hwss{} whose Verma modules are reducible, and hence have non-trivial singular vectors.  For $c = -2$, the corresponding dimensions are given by
\begin{equation}
h_{r,s} = \frac{\brac{2r-s}^2 - 1}{8} \qquad \text{($r,s \in \ZZ_+$).}
\end{equation}

\begin{table}
\begin{center}
\setlength{\extrarowheight}{4pt}
\begin{tabular}{|C|C|C|C|C|C|C|C}
\hline
0 & -\tfrac{1}{8} & 0 & \tfrac{3}{8} & 1 & \tfrac{15}{8} & 3 & \cdots \\[1mm]
\hline
1 & \tfrac{3}{8} & 0 & -\tfrac{1}{8} & 0 & \tfrac{3}{8} & 1 & \cdots \\[1mm]
\hline
3 & \tfrac{15}{8} & 1 & \tfrac{3}{8} & 0 & -\tfrac{1}{8} & 0 & \cdots \\[1mm]
\hline
6 & \tfrac{35}{8} & 3 & \tfrac{15}{8} & 1 & \tfrac{3}{8} & 0 & \cdots \\[1mm]
\hline
10 & \tfrac{63}{8} & 6 & \tfrac{35}{8} & 3 & \tfrac{15}{8} & 1 & \cdots \\[1mm]
\hline
\cdots & \cdots & \cdots & \cdots & \cdots & \cdots & \cdots & \ddots
\end{tabular}
\vspace{3mm}
\caption{A part of the extended Kac table for $c=-2$, listing the conformal dimensions $h_{r,s}$.  Here, $r$ increases downwards, and $s$ increases to the right, so the top-left-hand corner corresponds to the vacuum with $h_{1,1} = 0$.} \label{tabKacTable}
\end{center}
\end{table}

The central charge $c = -2$ occupies a special place in the history of \lcft{} as it was here that the presence of logarithmic singularities in correlation functions was first linked to the presence of non-trivial Jordan cells in the normal form of $L_0$ \cite{GurLog93}.  Indeed, one of the best understood examples of these theories, the \emph{triplet model} of Gaberdiel and Kausch \cite{GabRat96,GabLoc99}, has $c = -2$.  This model has an unusual extended chiral algebra that is generated by an energy-momentum tensor $\mathsf{T}$ and three (Virasoro) primary fields $\mathsf{W}^i$ (whence the appellation ``triplet'') of dimension $3$.  The \opes{} take the form (when normalised appropriately)
\begin{multline} \label{eqnTripletOPEs}
\func{\mathsf{W}^i}{z} \func{\mathsf{W}^j}{w} = \\
\kappa^{ij} \sqbrac{\frac{-2}{\brac{z-w}^6} + \frac{6 \func{\mathsf{T}}{w}}{\brac{z-w}^4} + \frac{3 \func{\partial \mathsf{T}}{w}}{\brac{z-w}^3} - \frac{\frac{3}{2} \func{\partial^2 \mathsf{T}}{w} - 8 \func{\normord{\mathsf{T} \mathsf{T}}}{w}}{\brac{z-w}^2} - \frac{\frac{1}{3} \func{\partial^3 \mathsf{T}}{w} - 8 \func{\normord{\partial \mathsf{T} \mathsf{T}}}{w}}{z-w}} \\
+ \sum_k f^{ij}_{\phantom{ij} k} \sqbrac{\frac{5 \func{\mathsf{W}^k}{w}}{\brac{z-w}^3} + \frac{\frac{5}{2} \func{\partial \mathsf{W}^k}{w}}{\brac{z-w}^2} + \frac{\frac{1}{5} \func{\partial^2 \mathsf{W}^k}{w} + \frac{12}{5} \func{\normord{\mathsf{T} \mathsf{W}^k}}{w}}{z-w}} + \ldots,
\end{multline}
where $\kappa^{ij}$ and $f^{ij}_{\phantom{ij} k}$ are the Killing form and structure constants of $\SLA{sl}{2}$ with respect to some (arbitrary) basis.  This algebra is not the enveloping algebra of any (finitely generated) Lie algebra, but it nevertheless admits an obvious triangular decomposition.  The notions of \hws{} and \hwm{} are therefore well-defined, and it turns out that only four irreducible \hwms{} appear in the triplet model.

The relevant irreducibles turn out to be generated by \hwss{} of conformal dimensions $h_{1,1} = 0$, $h_{2,1} = 1$, $h_{1,2} = -\tfrac{1}{8}$ and $h_{2,2} = \tfrac{3}{8}$, respectively \cite{GabRat96}.  Furthermore, there is a single state (up to normalisation) of dimensions $0$ and $-\tfrac{1}{8}$ in the corresponding modules, whereas the multiplicity of the states of dimensions $1$ and $\tfrac{3}{8}$ in their modules is two.  We mention for later reference that the fusion rules of the theory close on the dimension $0$ and $1$ modules, but that fusion among the dimension $-\tfrac{1}{8}$ and $\tfrac{3}{8}$ modules leads to two new modules.  These may be identified as indecomposable extensions of the dimension $0$ and $1$ modules and are responsible for the logarithmic nature of the triplet model \cite{GabRat96}.  The fusion rules close on this larger set of modules.

For now we wish to concentrate on the two \hwms{} of the triplet algebra whose dimensions are $0$ and $1$.  All the states of these modules then have integer conformal dimension.  This should remind the reader of our coset theory.  By the end of \secref{secCoset}, we had determined that our coset had precisely $m+1$ Virasoro \hwss{} of dimension $m \brac{m+1} / 2$.  Taking $m=0$ and $1$, we thereby recover just one state of conformal dimension $0$ but two states of conformal dimension $1$, just as in the triplet model.  Moreover, the three primary fields $\mathsf{W}^i$ which generate the triplet algebra correspond to three Virasoro \hwss{} in the vacuum triplet module.  This also accords with the \hws{} multiplicities which we have observed in our coset theory (take $m=2$ in the above), leading us to conjecture that this coset theory is nothing but the triplet model.

More precisely, we conjecture that our coset model \eqref{eqnCosetModel} has the same extended chiral symmetry algebra as the triplet model.  This is a reasonable hope as we know that the fields of the $\AKMA{sl}{2}_{-1/2}$ vacuum module are chiral, so the fields into which these decompose in the coset model will also be chiral.  Moreover, by \eqnref{eqnCosetChar1} the ``extended'' coset vacuum module has character
\begin{equation} \label{eqnCosetVacChar}
\sum_{n \in \ZZ} \ \sideset{}{'} \sum_{m = 2 \abs{n}}^{\infty} \frac{q^{m \brac{m+1} / 2} - q^{\brac{m+1} \brac{m+2} / 2}}{\qnum{q}{\infty}} z^{2 n} = \frac{q^{-1/8}}{\qnum{q}{\infty}} \sum_{m \in \ZZ} q^{\brac{4m+1}^2 / 8} \cdot \frac{z^{2m+1} - z^{-2m-1}}{z - z^{-1}},
\end{equation}
which coincides with the character of the triplet vacuum module \cite{FloMod96,KauCur95} (the $z$-dependence can be verified within the symplectic fermion framework).  This coincidence suggests that the coset theory admits the same chiral algebra as the triplet model, but proving this requires some non-trivial arguments and computations.  We turn to these next.

In principle, we should start calculating correlation functions in the coset theory.  This would then enable us to deduce \opes{} for the chiral fields, and we would expect to thereby uncover those of the triplet algebra (\eqnref{eqnTripletOPEs}).  However, determining correlators of coset theories is generally acknowledged to be quite hard --- the coset mechanism is not well-suited for such computations \cite{DiFCon97} --- and in most cases one has to resort to a free-field realisation.  Luckily, here we can proceed chirally by using the $\beta \gamma$ ghost system, itself a free-field realisation of the $\AKMA{sl}{2}_{-1/2}$-theory that we started with.  We will show in what follows that the algebra structure of the coset of the ghosts by its $\AKMA{u}{1}$-subtheory is completely fixed by its vacuum module structure.  This will then allow us to deduce the chiral algebra of the $\AKMA{sl}{2}_{-1/2}$-coset.\footnote{In fact, the algebra of the triplet model can be fixed using only the underlying vacuum representation.  However, this requires a long and convoluted analysis of associativity and null-vectors (reported in \cite{KauExt91} though not detailed there).  We propose studying instead the ghost coset algebra because the corresponding analysis is extremely simple.}  We remark that this approach is constructive --- the rather complicated form of the triplet algebra \opes{} (\eqnref{eqnTripletOPEs}) will be verified as a simple corollary of those of the ghost coset theory.

To begin, we recall that the vacuum module of the ghost theory is formed from the $k = -\tfrac{1}{2}$ $\AKMA{sl}{2}$-modules $\AffIrrMod{0}$ and $\AffIrrMod{1}$ \cite{RidSL208}.  The two zero-grade states of the latter module (see \figref{figu(1)Mults}) give rise to two $\alg{vir}$-\hwss{} in the coset theory of dimension $1$.  We will denote these \hwss{} by $\ket{\zeta}$ and $\ket{\eta}$, noting that they have $\SLA{sl}{2}$-weights $1$ and $-1$ (respectively).  Since such weights are conserved in \opes{}, we can immediately deduce from \figref{figu(1)Mults} and our knowledge of conformal dimensions in the coset theory that
\begin{subequations} \label{eqnSFOPEs1}
\begin{align}
\func{\zeta}{z} \func{\zeta}{w} &= \func{W^+}{w} \brac{z-w} + \ldots & \func{\eta}{z} \func{\zeta}{w} &= \frac{a}{\brac{z-w}^2} + b \func{\mathbb{T}}{w} + \ldots \\
\func{\zeta}{z} \func{\eta}{w} &= \frac{\mu a}{\brac{z-w}^2} + \mu b \func{\mathbb{T}}{w} + \ldots & \func{\eta}{z} \func{\eta}{w} &= \func{W^-}{w} \brac{z-w} + \ldots,
\end{align}
\end{subequations}
where $a$ and $b$ are unknown constants, $\mu = \pm 1$ describes whether $\zeta$ and $\eta$ are mutually bosonic or fermionic, and the $\func{W^{\pm}}{w}$ are Virasoro primaries of dimension $3$ and $\SLA{sl}{2}$-weight $\pm 2$.  It follows immediately that
\begin{equation}
\func{\mathbb{T}}{z} = b^{-1} \normord{\func{\eta}{z} \func{\zeta}{z}}.
\end{equation}

Assuming that $\zeta$ and $\eta$ are mutually bosonic ($\mu = 1$) leads to $\mathbb{T}$ having central charge $2$, whereas assuming that they are mutually fermionic ($\mu = -1$) gives $c=-2$.  With either choice, $\func{\mathbb{T}}{z} \func{\mathbb{T}}{w}$ has the wrong \ope{} for an energy-momentum tensor unless $a+b=0$.  This fixes $\zeta$ and $\eta$ to be mutually fermionic and the \opes{} \eqref{eqnSFOPEs1} require that in addition, $\zeta$ and $\eta$ are both mutually fermionic with respect to themselves.  Normalising so that $a=1$, the defining \opes{} become
\begin{subequations} \label{eqnSFOPEs}
\begin{align}
\func{\zeta}{z} \func{\zeta}{w} &= \func{W^+}{w} \brac{z-w} + \ldots & \func{\eta}{z} \func{\zeta}{w} &= \frac{1}{\brac{z-w}^2} - \func{\mathbb{T}}{w} + \ldots \\
\func{\zeta}{z} \func{\eta}{w} &= \frac{-1}{\brac{z-w}^2} + \func{\mathbb{T}}{w} + \ldots & \func{\eta}{z} \func{\eta}{w} &= \func{W^-}{w} \brac{z-w} + \ldots
\end{align}
\end{subequations}
We recognise these as the defining relations of the chiral algebra of (a subtheory of) symplectic fermions \cite{KauSym00}.

This proves that the $\AKMA{u}{1}$-coset of the $\beta \gamma$ ghost system admits the same chiral algebra as the theory of symplectic fermions.  To deduce from this the corresponding result for $\AKMA{sl}{2}_{-1/2}$ and the triplet model, we will show two things:  First, that the triplet algebra can be derived as a subalgebra of the chiral algebra of symplectic fermions, and second, that the fields generating the triplet algebra are expressible in terms of the coset representatives of the chiral fields of our $\AKMA{sl}{2}_{-1/2}$-theory.  In fact, this first requirement is already well known --- such a demonstration may be found for example in \cite{KauCur95} --- but we will nevertheless outline the derivation for completeness.

Indeed, the triplet algebra generators are now evident.  We define
\begin{equation} \label{eqnDefW+-}
\func{W^+}{z} = \normord{\func{\partial \zeta}{z} \func{\zeta}{z}} \qquad \text{and} \qquad \func{W^-}{z} = \normord{\func{\partial \eta}{z} \func{\eta}{z}}
\end{equation}
and check that these are indeed primary fields of dimension $3$ with respect to $\mathbb{T}$.  The \ope{} of $W^{\pm}$ with itself is regular, as expected, and we compute (with the help of \textsc{OPEdefs} \cite{ThiOPE91}) that
\begin{multline} \label{eqnCosetOPE1}
\func{W^{\pm}}{z} \func{W^{\mp}}{w} = \\
-\sqbrac{\frac{-2}{\brac{z-w}^6} + \frac{6 \func{\mathbb{T}}{w}}{\brac{z-w}^4} + \frac{3 \func{\partial \mathbb{T}}{w}}{\brac{z-w}^3} - \frac{\frac{3}{2} \func{\partial^2 \mathbb{T}}{w} - 8 \func{\normord{\mathbb{T} \mathbb{T}}}{w}}{\brac{z-w}^2} - \frac{\frac{1}{3} \func{\partial^3 \mathbb{T}}{w} - 8 \func{\normord{\partial \mathbb{T} \mathbb{T}}}{w}}{z-w}} \\
\mp \sqbrac{\frac{5 \func{W^0}{w}}{\brac{z-w}^3} + \frac{\frac{5}{2} \func{\partial W^0}{w}}{\brac{z-w}^2} + \frac{\frac{1}{5} \func{\partial^2 W^0}{w} + \frac{12}{5} \func{\normord{\mathbb{T} W^0}}{w}}{z-w}} + \ldots,
\end{multline}
where $\func{W^0}{w}$ is the third Virasoro primary of dimension $3$ (its $\SLA{sl}{2}$-weight vanishes), normalised as
\begin{equation} \label{eqnDefW0}
\func{W^0}{z} = \normord{\func{\partial \zeta}{z} \func{\eta}{z}} - \normord{\func{\zeta}{z} \partial \func{\eta}{z}}.
\end{equation}
We mention that $\func{W^0}{w}$ appears, along with $\func{\partial T}{w}$, in the first omitted term of the \opes{} \eqref{eqnSFOPEs} of $\zeta$ and $\eta$.  The remaining \opes{} of the triplet generators turn out to be
\begin{equation} \label{eqnCosetOPE2}
\func{W^0}{z} \func{W^{\pm}}{w} = \pm 2 \sqbrac{\frac{5 \func{W^{\pm}}{w}}{\brac{z-w}^3} + \frac{\frac{5}{2} \func{\partial W^{\pm}}{w}}{\brac{z-w}^2} + \frac{\frac{1}{5} \func{\partial^2 W^{\pm}}{w} + \frac{12}{5} \func{\normord{\mathbb{T} W^{\pm}}}{w}}{z-w}} + \ldots
\end{equation}
and
\begin{multline} \label{eqnCosetOPE3}
\func{W^0}{z} \func{W^0}{w} = \\
2 \sqbrac{\frac{-2}{\brac{z-w}^6} + \frac{6 \func{\mathbb{T}}{w}}{\brac{z-w}^4} + \frac{3 \func{\partial \mathbb{T}}{w}}{\brac{z-w}^3} - \frac{\frac{3}{2} \func{\partial^2 \mathbb{T}}{w} - 8 \func{\normord{\mathbb{T} \mathbb{T}}}{w}}{\brac{z-w}^2} - \frac{\frac{1}{3} \func{\partial^3 \mathbb{T}}{w} - 8 \func{\normord{\partial \mathbb{T} \mathbb{T}}}{w}}{z-w}} + \ldots
\end{multline}
Comparing Equations~\eqref{eqnCosetOPE1}, \eqref{eqnCosetOPE2} and \eqref{eqnCosetOPE3} with \eqnref{eqnTripletOPEs}, we conclude that $W^+$, $W^0$ and $W^-$ indeed realise the triplet algebra with the Killing form and structure constants of $\SLA{sl}{2}$ for the $\SLA{sl}{2 ; \RR}$-type basis $\set{e,h,f}$ of \secref{secOld} ($W^+$ corresponds to $e$, $W^0$ to $h$ and $W^-$ to $f$).

The triplet algebra therefore appears as a subalgebra of the chiral algebra of symplectic fermions.  We note that because the triplet generators $W^+$, $W^0$ and $W^-$ have dimension $3$ and $\SLA{sl}{2}$-weights $2$, $0$ and $-2$ respectively, they must be coset representatives of fields associated to the $\AKMA{sl}{2}$-module $\AffIrrMod{0}$.\footnote{This follows from the fact that the other possibility, $\AffIrrMod{1}$, has only odd $\SLA{sl}{2}$-weights.  We remark that in \secref{secAugment}, we will introduce additional $\AKMA{sl}{2}$-modules $\AffOthMod{0}$ and $\AffOthMod{1}$ whose $\SLA{sl}{2}$-weights are even and odd, respectively.  But the dimensions of the corresponding coset fields are never integers, so fields from these modules can be ruled out as well.}  The triplet generators are therefore coset representatives of chiral $\AKMA{sl}{2}$-fields, completing our proof that the coset theory $\AKMA{sl}{2}_{-1/2} / \AKMA{u}{1}$ admits the triplet algebra as an extended chiral algebra.

In fact, it is easy to see from the coset conformal dimensions that $W^+$ is the representative of $e$ and $W^-$ is the representative of $f$, up to some normalisation.  The field corresponding to $W^0$ requires more work, but a straight-forward computation verifies that the only $\AKMA{u}{1}$-primary field of dimension $3$ and $\SLA{sl}{2}$-weight $0$ which is also primary with respect to the coset energy-momentum tensor $\mathbb{T}$ is
\begin{equation}
\partial^2 h - 6 \normord{\partial h h} - 8 \normord{h h h} - 27 \normord{\partial f e} + 3 \normord{\partial e f} + 12 \normord{f h e}.
\end{equation}
This field therefore has coset representative $W^0$, again up to some normalisation.

So much for the chiral algebra.  Could it be that our coset theory \emph{is} just the triplet model?  We know that the $\AKMA{sl}{2}$-module $\AffIrrMod{1}$ decomposes in the coset theory into a module with character (compare \eqnref{eqnCosetVacChar})
\begin{equation} \label{eqnCosetCharI}
\sum_{n \in \ZZ + 1/2} \ \sideset{}{'} \sum_{m = 2 \abs{n}}^{\infty} \frac{q^{m \brac{m+1} / 2} - q^{\brac{m+1} \brac{m+2} / 2}}{\qnum{q}{\infty}} z^{2 n} = \frac{q^{-1/8}}{\qnum{q}{\infty}} \sum_{m \in \ZZ} q^{\brac{4m-1}^2 / 8} \cdot \frac{z^{2m} - z^{-2m}}{z - z^{-1}},
\end{equation}
which coincides with the character of the irreducible triplet module whose \hws{} has conformal dimension $1$ \cite{KauCur95}.  A nice corollary of the above chiral algebra identification is that we can now conclude that $\AffIrrMod{1}$ does indeed become this irreducible triplet module in the coset model, because the triplet algebra admits no other module with this character.  As for the other modules of $\AKMA{sl}{2}_{-1/2}$, we have already noted in \secref{secCoset} that these other modules are twisted versions of either $\AffIrrMod{0}$ or $\AffIrrMod{1}$ and so they yield identical coset modules.\footnote{In fact, the spectral flow generator does shift the set of $\SLA{sl}{2}$-weights by multiples of $k = -\tfrac{1}{2}$, so the coset modules would be distinguishable on this basis.  However, the triplet algebra cannot see this as it does not contain any element like $h_0$.}  The standard ``field identification'' now removes them from further consideration.

But we also have to wonder about the remaining modules of the triplet algebra.  What we have shown so far is that our coset theory consists of two irreducible triplet modules, both of which have states of integer conformal dimension.  This is not true for the other irreducible triplet modules (recall that their \hwss{} had conformal dimensions $-\tfrac{1}{8}$ and $\tfrac{3}{8}$), so the triplet model cannot be our coset theory.  Furthermore, we have already noted that the triplet model is a \lcft{}, so there are additional indecomposable (but reducible) modules in the theory.

The coset theory we have constructed above is therefore definitely not the triplet model, but this reasoning suggests that we should identify our coset theory with a part of the triplet model.\footnote{But is this \emph{partial} coset theory well-defined in itself?  The answer appears to be ``no''.  In particular, it has no modular invariant.  This is very interesting to note, especially when one recalls that the standard method for analysing cosets of rational theories is to first establish the modular properties of the coset characters and then work one's way back \cite{DiFCon97}.  A more interesting question is then whether the non-logarithmic parent theory $\AKMA{sl}{2}_{-1/2}$ is well-defined!  We hope to report on this in the future.}  One possible conclusion is then that our coset theory is not complete because the theory we started with was too small.  We will therefore return to the fractional level theory $\AKMA{sl}{2}_{-1/2}$, seeking a natural augmentation of the spectrum.  Since the triplet model is logarithmic, we expect that this augmentation will also lead to a \lcft{}.

\section{Augmenting $\AKMA{sl}{2}_{-1/2}$} \label{secAugment}

At first sight, it seems unreasonable to try to augment the fractional level $\AKMA{sl}{2}_{-1/2}$ theory.  It is well known \cite{LesSU202,RidSL208} that the structure of the irreducible vacuum module selects the four admissible modules of Kac and Wakimoto as the only \hwms{} allowed in the theory.  These are all irreducible, and all four are required if we insist upon modular invariance \cite{KacMod88b}.  However, these \hwms{} do not close under fusion or conjugation, rather they generate infinite sequences of irreducible modules which are not highest weight (\secref{secOld}).  It is therefore germane to wonder if there are not other non-highest weight ``admissible'' modules (in the sense that they are likewise selected by the structure of the irreducible vacuum module) which we have not yet considered.

A strong clue as to where we should look for such additional admissible modules is contained in Gaberdiel's work \cite{GabFus01} on $\AKMA{sl}{2}_{-4/3}$.  There, it was found that the admissible \hwms{} and their spectral flows do not close under fusion but also generate modules whose zero-grade subspace is neither highest nor lowest weight for $\SLA{sl}{2}$.  Such modules are not generated by fusion in our $\AKMA{sl}{2}_{-1/2}$ model, but they do form natural candidates for augmenting the theory.

Consider therefore the (non-trivial) vanishing singular vector of the vacuum $\AKMA{sl}{2}$-module at $k = -\tfrac{1}{2}$.  This has $\SLA{sl}{2}$-weight $4$ and conformal dimension $4$ and is given explicitly (up to normalisation) by
\begin{equation}
\brac{156 e_{-3} e_{-1} - 71 e_{-2}^2 + 44 e_{-2} h_{-1} e_{-1} - 52 h_{-2} e_{-1}^2 + 16 f_{-1} e_{-1}^3 - 4 h_{-1}^2 e_{-1}^2} \ket{0}.
\end{equation}
The corresponding (vanishing) field is therefore
\begin{equation}
78 \normord{\partial^2 e e} - 71 \normord{\partial e \partial e} + 44 \normord{\partial e h e} - 52 \normord{\partial h e e} + 16 \normord{f e e e} - 4 \normord{h h e e} = 0.
\end{equation}
We have omitted the customary explicit coordinate dependence of this field for conciseness.  For computations, it is more convenient to consider instead its weight $0$ descendant of the same dimension (which therefore also vanishes):
\begin{multline}
64 \normord{e e f f} + 16 \normord{e h h f} - 136 \normord{e h \partial f} + 128 \normord{e \partial h f} - 12 \normord{e \partial^2 f} - 8 \normord{h h h h} \\
+ 200 \normord{\partial e h f} - 108 \normord{\partial e \partial f} + 8 \normord{\partial h h h} - 38 \normord{\partial h \partial h} + 156 \normord{\partial^2 e f} + 24 \normord{\partial^2 h h} - \partial^3 h = 0.
\end{multline}
We expand this (chiral) field into modes and let the zero-mode act on a state $\ket{v_m}$ of $\SLA{sl}{2}$-weight $m$ which we assume is annihilated by each $e_j$, $h_j$ and $f_j$ with $j \geqslant 1$.  Such states $\ket{v_m}$ are not necessarily $\AKMA{sl}{2}$-\hwss{} and have sometimes been referred to as \emph{relaxed} \hwss{} in the literature \cite{FeiEqu98,SemEmb97}.  The resulting constraint is
\begin{equation} \label{eqnRelHWSCon}
\brac{64 f_0^2 e_0^2 + 16 f_0 h_0^2 e_0 - 192 f_0 h_0 e_0 + 180 f_0 e_0 - 8 h_0^4 - 8 h_0^3 + 10 h_0^2 + 6 h_0} \ket{v_m} = 0.
\end{equation}

If $\ket{v_m}$ is indeed an $\AKMA{sl}{2}$-\hws{}, this constraint restricts $m$ to be one of the admissible weights $0$, $1$, $-\tfrac{1}{2}$ or $-\tfrac{3}{2}$.  If not, we have to work a little harder.  Our conventions for $\SLA{sl}{2}$-modules which are not highest weight are summarised in \appref{appSL2}.  In particular, \eqnref{eqnDefAlpha} defines constants $\alpha_m$ by
\begin{equation}
\ket{v_{m + 2}} = e \ket{v_m} \qquad \text{and} \qquad f \ket{v_m} = \alpha_{m} \ket{v_{m - 2}},
\end{equation}
and \eqnref{eqnAlphaRel} shows that they are related by
\begin{equation} \label{eqnAlphRel}
\alpha_{m+2} = \alpha_{m} + m.
\end{equation}
The constraint \eqref{eqnRelHWSCon} therefore becomes
\begin{equation}
64 \alpha_{m + 4} \alpha_{m + 2} + \brac{16 \brac{m + 2}^2 - 192 \brac{m + 2} + 180} \alpha_{m + 2} - 8 m^4 - 8 m^3 + 10 m^2 + 6 m = 0,
\end{equation}
and substituting \eqnref{eqnAlphRel} for $\alpha_{m + 4}$ allows us to solve this explicitly.  The result is that there are only two possibilities:
\begin{equation} \label{eqnPossibilities}
\alpha_{m} = -\frac{m \brac{m + 1}}{2} \qquad \text{or} \qquad \alpha_{m} = \frac{\brac{2 m - 1} \brac{2 m - 3}}{16}.
\end{equation}

As we already know the \hwss{}, we may suppose that $\ket{v_m}$ is not such a state.  If both $\alpha_{m}$ and $\alpha_{m + 2}$ are given by the first possibility of \eqnref{eqnPossibilities}, then substituting into \eqnref{eqnAlphRel} implies that $m = -1$, hence that $\alpha_{m} = 0$.  We thereby deduce that in this case $\ket{v_{-1}}$ is the non-highest weight zero-grade state of the $\AKMA{sl}{2}$-module $\AffIrrMod{1}$.  If $\alpha_{m}$ and $\alpha_{m + 2}$ are given by different possibilities in \eqnref{eqnPossibilities} then we also get constraints on the weights $m$.  However, in this case one finds that $m \notin \RR$ which is impossible if we want to equip the corresponding module with a (non-zero) hermitian form that respects $h_0^{\dag} = h_0$ (\appref{appSL2}).

It remains then to study when both $\alpha_{m}$ and $\alpha_{m + 2}$ are given by the second possibility of \eqnref{eqnPossibilities}, and it turns out that there is now no constraint upon $m$.  To understand what \eqnref{eqnPossibilities} means in this case, we note that it is equivalent to the eigenvalue of the $\SLA{sl}{2}$-Casimir $Q = \tfrac{1}{2} h^2 - ef - fe$ being
\begin{equation}
Q_{m} = \frac{1}{2} m^2 - m - 2 \alpha_{m} = -\frac{3}{8}.
\end{equation}
Recalling that $\func{T}{z}$ is closely related to $Q$ (see \eqnref{eqnDefT}), this means that a relaxed \hws{} $\ket{v_m}$ is admissible --- which we henceforth (re)define to mean not forbidden by the structure of the irreducible vacuum module --- if its conformal dimension is $\tfrac{1}{3} Q_{m} = -\frac{1}{8}$.  This covers all the zero-grade states of the (twisted) $\AKMA{sl}{2}$-modules $\AffIrrMod{-1/2}$ and $\AffIrrMod{-3/2}$ and their conjugates.

However, it also allows for many more possibilities.  In particular, given an arbitrary weight $m$, we can construct relaxed \hwss{} $\ket{v_m}$ with conformal dimension $-\frac{1}{8}$ by choosing
\begin{equation} \label{eqnAlphaLambda}
\alpha_{m} = \frac{\brac{2 m - 1} \brac{2 m - 3}}{16}.
\end{equation}
These states are therefore admissible, and constitute a zero-grade subspace which has neither highest nor lowest weight states (as an $\SLA{sl}{2}$-module).\footnote{Actually, here we must stipulate that $m \notin \ZZ + \tfrac{1}{2}$, for the alternative leads to $\alpha_m = 0$ for $m = \tfrac{1}{2}$ or $\tfrac{3}{2}$, indicating a lowest weight state.  Nevertheless, zero-grade states with $m \in \ZZ + \tfrac{1}{2}$ are not excluded from being admissible.}  This analysis, similar to that reported in \cite{GabFus01,LesLog04}, therefore indicates that there should be further modules, built from these zero-grade subspaces, which are not forbidden in an $\AKMA{sl}{2}_{-1/2}$-theory.  We now ask ourselves how we can decide which of these modules we should choose to augment our current theory by, so as to obtain the triplet model as a coset theory.

The answer to our question lies in determining the conformal dimensions of the coset theory states corresponding to these new zero-grade states.  Since the relaxed \hwss{} are assumed to be annihilated by the $h_j$ with $j > 0$, \eqnref{eqnDefCosetL0} gives the conformal dimension as
\begin{equation}
\frac{1}{2} m^2 - \frac{1}{8}.
\end{equation}
Note that for $m$ integral, this formula precisely reproduces the dimensions $h_{r,2}$ of the extended Kac table, \tabref{tabKacTable}, with $r = \abs{m} + 1$.  It follows that a \emph{relaxed} highest weight $\AKMA{sl}{2}$-module whose zero-grade subspace consists of relaxed \hwss{} of conformal dimension $-\frac{1}{8}$ and $\SLA{sl}{2}$-weights in $2 \ZZ$ will decompose in the coset model into Virasoro modules with \hwss{} of dimensions $-\frac{1}{8}$, $\frac{15}{8}$, $\frac{63}{8}$ and so on.  Similarly, when the zero-grade subspace has $\SLA{sl}{2}$-weights in $2 \ZZ + 1$, the Virasoro \hwss{} will have dimensions $\frac{3}{8}$, $\frac{35}{8}$, $\frac{99}{8}$ and so on.  This is obviously suggestive for the two irreducible triplet algebra representations discussed at the end of \secref{secTriplet}.  Indeed, we mention that this decomposition gives a single Virasoro-\hws{} of dimension $-\frac{1}{8}$ but two of dimension $\frac{3}{8}$, just as one finds in these triplet algebra representations.

Let us therefore consider the augmentation of the $\AKMA{sl}{2}_{-1/2}$-theory by the integer weight modules described above (and their twisted versions under the spectral flow).  We will refer to the theory generated by this augmentation as the $\AKMA{sl}{2}_{-1/2}^{\text{aug.}}$-theory.  In the language of \cite{LesLog04}, this is a \emph{lift} of the original $\AKMA{sl}{2}_{-1/2}$-theory, though we emphasise that the modules we are augmenting by were not considered in that article.

We want to be precise about the nature of these new $\AKMA{sl}{2}$-modules, so let us first consider the infinite-dimensional $\SLA{sl}{2}$-representations formed by their zero-grade subspaces.  These are defined by
\begin{equation} \label{eqnSL2PrincipalRep}
e \ket{v_m} = \ket{v_{m + 2}}, \qquad h \ket{v_m} = m \ket{v_m} \qquad \text{and} \qquad f \ket{v_m} = \frac{\brac{2 m - 1} \brac{2 m - 3}}{16} \ket{v_{m - 2}},
\end{equation}
for $m$ either all even or all odd.  Both have the required Casimir eigenvalue $Q_m = -\tfrac{3}{8}$.  Moreover, they lift to representations of the positive Borel subalgebra of $\AKMA{sl}{2}$ by imposing $e_j \ket{v_m} = h_j \ket{v_m} = f_j \ket{v_m} = 0$ for $j \geqslant 1$ (and declaring that $k = -\tfrac{1}{2}$), and thence to induced modules of $\AKMA{sl}{2}$ itself.  We denote these induced modules by $\AffIndMod{0}$ or $\AffIndMod{1}$ according to whether $m$ runs over the even or odd integers respectively.  They are examples of relaxed Verma modules in the language of \cite{FeiEqu98,SemEmb97}.  Because the $e_j$, $h_j$ and $f_j$ with $j \leqslant -1$ act freely, the characters of the $\AffIndMod{\lambda}$ are easily computed to be
\begin{align} \label{eqnCharIndSL2Mods}
\ch{\AffIndMod{\lambda}}{z ; q} &= \frac{q^{-1/8}}{\displaystyle \prod_{i=1}^{\infty} \brac{1 - z^{-2} q^i} \brac{1 - q^i} \brac{1 - z^2 q^i}} \sum_{n \in \ZZ + \lambda / 2} z^{2n} = \frac{q^{-1/8}}{\qnum{q}{\infty}} \sum_{n \in \ZZ + \lambda / 2} \ \sum_{i,j=0}^{\infty} \frac{q^{i+j}}{\qnum{q}{i} \qnum{q}{j}} z^{2 \brac{n+j-i}} \notag \\
&= \frac{q^{-1/8}}{\qnum{q}{\infty}} \sum_{i=0}^{\infty} \frac{q^i}{\qnum{q}{i}} \sum_{j=0}^{\infty} \frac{q^j}{\qnum{q}{j}} \sum_{n \in \ZZ + \lambda / 2} z^{2n} = \frac{q^{-1/8}}{\qnum{q}{\infty}^3} \sum_{n \in \ZZ + \lambda / 2} z^{2n}.
\end{align}
Such induced modules are universal for relaxed modules in the same way that Verma modules are for \hwms{}.

As with genuine Verma modules, we can ask if these relaxed Verma modules are themselves irreducible.  It is easy to show inductively that any proper submodule of the $\AffIndMod{\lambda}$ must be generated by \hwss{} (relaxed or genuine), and we have seen that the vanishing vacuum singular vector forbids these from appearing in the theory unless they have conformal dimension $0$, $\tfrac{1}{2}$ or $-\tfrac{1}{8}$.  As these dimensions are impossible for generators of proper submodules of $\AffIndMod{0}$ and $\AffIndMod{1}$ (recall that the zero-grade states of these modules have dimension $-\tfrac{1}{8}$), any such proper submodules must be set to zero.  It follows that we should not try to augment our theory by the relaxed Verma modules $\AffIndMod{0}$ or $\AffIndMod{1}$, but rather by their irreducible quotients.  We will denote the irreducible quotients of $\AffIndMod{0}$ or $\AffIndMod{1}$ by $\AffOthMod{0}$ and $\AffOthMod{1}$ (respectively).  That these irreducibles are indeed proper quotients of the relaxed Verma modules is demonstrated by the states at grade $1$ of the form
\begin{equation} \label{eqnETypeSVs1}
\frac{\brac{2 m - 1} \brac{2 m + 3}}{16} e_{-1} \ket{v_{m - 2}} - \frac{2 m + 3}{4} h_{-1} \ket{v_m} + f_{-1} \ket{v_{m + 2}},
\end{equation}
which may be checked to be relaxed \hwss{} (in fact this is true for all $m$, not just for $m \in \ZZ$).  These states therefore generate proper (relaxed) submodules of the $\AffIndMod{\lambda}$ and so are set to zero in the corresponding irreducibles $\AffOthMod{\lambda}$.

While there does exist a detailed structure theory of relaxed Verma modules for $\AKMA{sl}{2}$ \cite{FeiEqu98,SemEmb97}, it is perhaps easier to obtain the characters of the irreducible $\AKMA{sl}{2}$-modules $\AffOthMod{0}$ and $\AffOthMod{1}$ by using the free field realisation of the $\beta \gamma$ ghost system.  This does depend upon the fact that the simple current module $\AffIrrMod{1}$ of $\AKMA{sl}{2}_{-1/2}$ remains a simple current in $\AKMA{sl}{2}_{-1/2}^{\text{aug.}}$, more precisely that
\begin{equation} \label{eqnLEFusion}
\AffIrrMod{1} \fuse \AffOthMod{0} = \AffOthMod{1} \qquad \text{and} \qquad \AffIrrMod{1} \fuse \AffOthMod{1} = \AffOthMod{0}.
\end{equation}
The computation of these fusion rules will be deferred to \cite{RidFus10}, along with the rest of the fusion rules of $\AKMA{sl}{2}_{-1/2}^{\text{aug.}}$.  Taking the above rules as given however, we have the natural interpretation that the two irreducible $\AKMA{sl}{2}$-modules $\AffOthMod{0}$ and $\AffOthMod{1}$ combine to form a single irreducible module for the $\beta \gamma$ ghost algebra.  This latter module will therefore have a zero-grade subspace with one state $\ket{v_m}$ for each integral weight $m \in \ZZ$.

It is easy to work out the character of the corresponding relaxed Verma module for the $\beta \gamma$ ghosts.  The same computation as in \eqnref{eqnCharIndSL2Mods} gives
\begin{equation}
\frac{q^{-1/8}}{\prod_{i=1}^{\infty} \brac{1 - z^{-1} q^i} \brac{1 - z q^i}} \sum_{m \in \ZZ} z^{m} = \frac{q^{-1/8}}{\qnum{q}{\infty}^2} \sum_{m \in \ZZ} z^{m}.
\end{equation}
Comparing with the sum of the characters of the relaxed Verma modules of $\AKMA{sl}{2}$ given in \eqnref{eqnCharIndSL2Mods}, we see that the character of this $\beta \gamma$-module is the same up to multiplication by the simple factor
\begin{equation} \label{eqnExpansion}
\qnum{q}{\infty} = 1 - q - q^2 + q^5 + q^7 - q^{12} - q^{15} + q^{22} + q^{26} - \ldots
\end{equation}
The multiplicities of the $\beta \gamma$-module are therefore smaller in general, so it seems plausible that this $\beta \gamma$-Verma module is in fact irreducible.  This is to be expected based on status of the $\beta \gamma$ ghost system as a free field theory (or as a simple current extension \cite{RidSU206,RidMin07}).  Indeed, we can prove this irreducibility by looking at the inner products of states in the $\beta \gamma$-module.\footnote{We could also verify explicitly that \eqnDref{eqnETypeSVs1}{eqnETypeSVs2} (below) vanish identically upon rewriting them using only the $\beta \gamma$ modes.}  To wit, this module has a Poincar\'{e}-Birkhoff-Witt basis of the form \cite{MooLie95}
\begin{equation}
\set{\beta_{-i_1} \cdots \beta_{-i_r} \gamma_{-j_1} \cdots \gamma_{-j_s} \ket{v_m} \st i_1 \geqslant \cdots \geqslant i_r \geqslant 1, \ j_1 \geqslant \cdots \geqslant j_s \geqslant 1, \ m \in \ZZ},
\end{equation}
which is easily checked to constitute an orthogonal basis using the commutation relations of \eqnref{eqnGhostAdjComm}.  The norms of these basis elements have the form $\brac{-1}^{j_1 + \ldots j_s} \braket{v_m}{v_m}$, and $\braket{v_m}{v_m} \neq 0$ by \eqnref{eqnNormRel} (as $\alpha_{m} \neq 0$ for $m \in \ZZ$ by \eqnref{eqnAlphaLambda}).  The module therefore has no null states, so its irreducibility follows from standard arguments.

We pause briefly to mention that the expansion \eqref{eqnExpansion} strongly suggests that the relaxed Verma modules $\AffIndMod{\lambda}$ at $k = -\tfrac{1}{2}$ have the same ``braided'' submodule structure as the ``admissible'' Verma modules.  More precisely, the above expansion is consistent with the picture that there are two independent submodules, generated by states of grade $1$ and $2$ respectively, whose intersection is the sum of two independent submodules, generated by states of grade $5$ and $7$ respectively, and so on.  In fact, the submodule structure of relaxed Verma modules was elucidated in \cite{FeiEqu98} and is in agreement with this observation.  The submodule generators at grade $1$ have already been given in \eqnref{eqnETypeSVs1} and those at grade $2$ are easily found to be
\begin{multline} \label{eqnETypeSVs2}
\frac{\brac{2m-7} \brac{2m-3} \brac{2m+1} \brac{2m+5}}{256} e_{-1}^2 \ket{v_{m-4}} - \frac{\brac{2m-3} \brac{2m+1} \brac{2m+5}}{32} \brac{h_{-1} e_{-1} - e_{-2}} \ket{v_{m-2}} \\
+ \frac{\brac{2m+1} \brac{2m+5}}{16} \brac{h_{-1}^2 + 2 f_{-1} e_{-1} - h_{-2}} \ket{v_m} - \frac{2m+5}{2} \brac{f_{-1} h_{-1} - f_{-2}} \ket{v_{m + 2}} + f_{-1}^2 \ket{v_{m + 4}}
\end{multline}
(and checked to be independent).  We note that the coefficients of this formula have a surprisingly regular form.

The conclusion of this exercise is then that the irreducible $\AKMA{sl}{2}$-modules $\AffOthMod{\lambda}$ have characters of the form
\begin{equation} \label{eqnCharEType}
\ch{\AffOthMod{\lambda}}{z ; q} = \frac{q^{-1/8}}{\qnum{q}{\infty}^2} \sum_{n \in \ZZ + \lambda / 2} z^{2n},
\end{equation}
so that their weight space multiplicities do not depend upon the $\SLA{sl}{2}$-weight.  We can now play the same game as in \secref{secCoset} to decompose $\AffOthMod{0}$ and $\AffOthMod{1}$ into $c = -2$ Virasoro modules in the coset theory.  In fact, decomposing the subspace of $\SLA{sl}{2}$-weight $2n \in \ZZ$ states into $\AKMA{u}{1}$-modules (recalling the latter's character from \eqnref{eqnU(1)Char}) corresponds to the simple character decomposition
\begin{equation} \label{eqnCosetChar2}
\frac{q^{-1/8}}{\qnum{q}{\infty}^2} = \frac{q^{-2 n^2}}{\qnum{q}{\infty}} \frac{q^{\brac{16 n^2 - 1}/8}}{\qnum{q}{\infty}} = \frac{q^{-2 n^2}}{\qnum{q}{\infty}} \frac{q^{h_{2 \abs{n} + 1,2}}}{\qnum{q}{\infty}}.
\end{equation}
This latter factor, $q^{h_{2 \abs{n} + 1,2}} / \qnum{q}{\infty}$, is therefore the character of the $c=-2$ Virasoro module corresponding to the weight $2n$ subspace.

In contrast to the constant $\SLA{sl}{2}$-weight characters of \eqnref{eqnCosetChar1}, this character does not uniquely determine a Virasoro module.  For example, it is obviously the character of the Virasoro Verma module with highest weight $h_{2 \abs{n} + 1,2}$.  But one can check that this character is shared by the (completely reducible) direct sum
\begin{equation} \label{eqnMoreMods}
\sideset{}{'} \bigoplus_{m = 2 \abs{n}}^{\infty} \IrrMod{\brac{2m-1} \brac{2m+1} / 8},
\end{equation}
where the primed summation means again that the sum index $m$ increases by $2$.  Of course there are many other possibilities as well.  Comparing with \eqnref{eqnSumModules}, we might suspect that \eqref{eqnMoreMods} is indeed the correct module structure for the weight $2n$ subspace of the coset decomposition of the $\AffOthMod{\lambda}$.  By the way of evidence for this, we note that the first singular vector in the Verma module with $h = h_{1,2} = -\tfrac{1}{8}$ (and $\SLA{sl}{2}$-weight $0$),
\begin{equation}
\Bigl( \mathbb{L}_{-1}^2 - \frac{1}{2} \mathbb{L}_{-2} \Bigr) \ket{v_0} = \Bigl( L_{-1}^2 - \frac{1}{2} L_{-2} - \frac{1}{4} h_{-1}^2 \Bigr) \ket{v_0},
\end{equation}
has zero norm in the module $\AffIndMod{0}$, hence must vanish in $\AffOthMod{0}$.  This singular vector is therefore zero in the coset theory, ruling out the Verma module possibility.

This does not prove that the coset module decomposes as in \eqref{eqnMoreMods}.  However, we do not need such a result because we have already determined that the chiral algebra of the coset theory is the triplet algebra.  By massaging the above characters as in \eqnDref{eqnCosetVacChar}{eqnCosetCharI}, we conclude that $\AffOthMod{0}$ and $\AffOthMod{1}$ give rise to coset modules with respective characters
\begin{equation}
\frac{q^{-1/8}}{\qnum{q}{\infty}} \sum_{m \in \ZZ} q^{\brac{4m}^2 / 8} \cdot \frac{z^{2m+1} - z^{-2m-1}}{z-z^{-1}} \qquad \text{and} \qquad \frac{q^{-1/8}}{\qnum{q}{\infty}} \sum_{m \in \ZZ} q^{\brac{4m-2}^2 / 8} \cdot \frac{z^{2m} - z^{-2m}}{z-z^{-1}}.
\end{equation}
Since these coincide with the irreducible triplet module characters with \hwss{} of dimensions $-\tfrac{1}{8}$ and $\tfrac{3}{8}$ respectively \cite{KauCur95}, we can conclude that the coset modules described above are precisely these triplet modules.

What we have thus shown is that augmenting $\AKMA{sl}{2}_{-1/2}$ by the admissible irreducible modules $\AffOthMod{0}$ and $\AffOthMod{1}$ (and their twisted versions under spectral flow) leads to a $\AKMA{u}{1}$-coset theory which contains \emph{all} of the irreducible modules of the triplet model.  It only remains to show that this augmentation also generates indecomposable modules corresponding to those of the triplet model, thus identifying the coset theory as the triplet model.  We will accomplish this in a sequel \cite{RidFus10} by analysing the fusion rules of our augmented theory $\AKMA{sl}{2}_{-1/2}^{\text{aug.}}$ in detail.

\section{Discussion and Conclusions} \label{secConc}

In the preceding sections, we have identified the coset \cft{} obtained from $\AKMA{sl}{2}_{-1/2}$ and its $\AKMA{u}{1}$-subtheory.  While this coset resembles the $\ZZ_k$-parafermion theories of Zamolodchikov and Fateev \cite{ZamNon85} in form, we see little in the way of structural resemblance.  Instead, we find that this fractional analogue of the parafermions forms (a part of) the triplet model of Gaberdiel and Kausch \cite{GabRat96}.\footnote{Actually, this lack of parafermionic behaviour is probably due to the fact that the ``integral part'' \cite[Sec.~18.6]{DiFCon97} of the level $k = -\tfrac{1}{2}$ is $1$.  It would be very interesting to check for parafermionic behaviour at more general levels.  For example, $k = \tfrac{1}{2}$ has integral part equal to $3$.}  More precisely, we have shown that the $\AKMA{sl}{2}_{-1/2}$-theory detailed in \cite{RidSL208} yields two of the four irreducible triplet modules under this identification, and moreover that the coset theory admits the triplet algebra as an extended chiral symmetry algebra.

This left us in an interesting situation in which the coset theory was not modular invariant despite the invariance of the parent $\AKMA{sl}{2}_{-1/2}$-theory.  Of course, as we explained in \secref{secOld}, the modular invariance of the latter theory is quite a subtle affair, so perhaps it is not surprising that mismatches such as this can occur.  Nevertheless, it led to the realisation that the consistency of our coset theory was far from satisfactory (at least on a torus).  Such a situation is not unheard of, and generally one adds in some sort of twisted modules to restore modular invariance to the coset.  Here, the resolution was the same and of course previous studies of the triplet model tell us exactly which additional modules were required.  However, we took this one step further by insisting that these new modules should be obtainable from the $\AKMA{sl}{2}_{-1/2}$-theory.  In other words, we reverse-engineered our original $\AKMA{sl}{2}_{-1/2}$-theory, \emph{augmenting} it with additional modules, so as to guarantee the modular invariance of the coset theory.

Following this philosophy, the two irreducible triplet algebra modules which were missing from the coset theory were discovered to be obtainable from $\AKMA{sl}{2}$-modules, albeit from a class of modules which were not originally considered, the so-called relaxed \hwms{}.  Specifically, we found two modules, $\AffOthMod{0}$ and $\AffOthMod{1}$, which were the only (relaxed) \hwms{} to yield the missing triplet modules.  Moreover, both $\AffOthMod{0}$ and $\AffOthMod{1}$ were shown to be admissible in the sense that they are not forbidden by the structure of the vacuum module (mathematically, they are representations of the corresponding \emph{vertex algebra}).  This led to a proposed augmentation of the original $\AKMA{sl}{2}_{-1/2}$-theory by $\AffOthMod{0}$, $\AffOthMod{1}$ and their images under the spectral flow.  We have denoted the corresponding augmented theory by $\AKMA{sl}{2}_{-1/2}^{\text{aug.}}$.

In contrast to $\AKMA{sl}{2}_{-1/2}$, we do not expect to have identified the full spectrum of $\AKMA{sl}{2}_{-1/2}^{\text{aug.}}$.  Indeed, the additional two irreducible triplet modules which necessitated this augmentation are well-known to fuse into modules which are reducible but indecomposable \cite{GabRat96}.  It is these indecomposables which are responsible for the logarithmic nature of the triplet model \cft{}.  By now it should not be surprising to learn that the same turns out to be true for fusions of $\AffOthMod{0}$ and $\AffOthMod{1}$ (this will be addressed in a sequel \cite{RidFus10}).  The augmented theory $\AKMA{sl}{2}_{-1/2}^{\text{aug.}}$ is therefore likewise a \lcft{}.

\subsection{Comparison with \cite{LesLog04}}

A logarithmic version of the $\AKMA{sl}{2}_{-1/2}$-theory was previously proposed in \cite{LesLog04} where it was referred to as a logarithmic \emph{lift}.  Actually, the authors of this paper proposed two different lifts based on different versions of the free field realisation that they relied upon.  This realisation utilised two fermionic fields $\eta$ and $\partial \xi$ to generate the chiral algebra and the two proposed lifts corresponded to whether they extended this algebra with formal antiderivatives of one or both of these fields.

Extending with one formal antiderivative led them to the construction of two relaxed \hwms{} (for $\AKMA{sl}{2}$) similar to our $\AffOthMod{0}$ and $\AffOthMod{1}$, but with weights belonging to $2 \ZZ + \tfrac{1}{2}$ and $2 \ZZ + \tfrac{3}{2}$ respectively.  Neither module was irreducible --- in accordance with \eqnref{eqnSL2PrincipalRep}, the zero-grade states with $\SLA{sl}{2}$-weights $\tfrac{1}{2}$ and $\tfrac{3}{2}$ are lowest weight states for $\SLA{sl}{2}$ which thereby generate proper $\AKMA{sl}{2}$-submodules.  We will therefore denote these indecomposable modules of \cite{LesLog04} by $\AffOthMod{1/2}^-$ and $\AffOthMod{3/2}^-$ (the ``$-$'' superscript indicating the presence of a lowest weight state with the weight given in the subscript).

It now appears that this particular extended free field realisation amounts to another augmentation of $\AKMA{sl}{2}_{-1/2}$, different to that which we have proposed here.  Furthermore, the authors of \cite{LesLog04} claim that their augmentation is closed under fusion (without any justification) and leads to no logarithmic phenomena, the complete opposite to what we expect for our augmentation.  But we think it important to point out that their augmentation modules $\AffOthMod{1/2}^-$ and $\AffOthMod{3/2}^-$ are neither self-conjugate nor conjugate to one another --- the conjugates would be $\AffOthMod{-1/2}^+$ and $\AffOthMod{-3/2}^+$ respectively (in hopefully obvious notation) --- in contrast to $\AffOthMod{0}$ and $\AffOthMod{1}$.  We therefore conclude that even if the augmentation by $\AffOthMod{1/2}^-$ and $\AffOthMod{3/2}^-$ closes under fusion, it is automatically unsatisfactory as a candidate for a \cft{} because many of its fields will have no conjugate and so must decouple 
from all correlation functions.  The only recourse is to augment further by the conjugate modules $\AffOthMod{-1/2}^+$ and $\AffOthMod{-3/2}^+$.  We strongly suspect that this, more consistent, augmentation will not close under fusion.

We speculate that the reason why $\AffOthMod{-1/2}^+$ and $\AffOthMod{-3/2}^+$ were not uncovered in \cite{LesLog04} was because the chiral algebra was only extended by the single formal antiderivative $\xi$.  This clearly breaks the symmetry between $\partial \xi$ and $\eta$, so it seems reasonable to suppose that extending instead by the formal antiderivative of $\eta$ would lead to the conjugate modules $\AffOthMod{-1/2}^+$ and $\AffOthMod{-3/2}^+$, rather than $\AffOthMod{1/2}^-$ and $\AffOthMod{3/2}^-$ (after all, $\eta$ is the conjugate field to $\partial \xi$ in their setup).  If this is so, then we could conclude that the consistency of the augmented theory requires extending by \emph{both} formal antiderivatives.

Happily, the authors of \cite{LesLog04} also analysed this possibility.  Using symplectic fermions and a free boson, they proposed two $\AKMA{sl}{2}$-modules with logarithmic structures (non-diagonalisability of $L_0$).  These structures were not however derived from any underlying insight into augmentations of $\AKMA{sl}{2}_{-1/2}$, but were rather determined directly as a consequence of the known logarithmic structures of symplectic fermion modules \cite{KauSym00}.  Given however that symplectic fermions are intimately related to the triplet model, it therefore seems likely that the logarithmic structures pictured in \cite{LesLog04} will turn out to be similar or identical to those (which we expect to find) in our proposed augmentation by $\AffOthMod{0}$ and $\AffOthMod{1}$.  We refer to our forthcoming article \cite{RidFus10} for a more detailed discussion of this point.

Even if this turns out to be the case, there are compelling reasons to continue investigating the augmentation $\AKMA{sl}{2}_{-1/2}^{\text{aug.}}$ which we have proposed in this article.  First, it is logically motivated through studying the consistency of the theory and its cosets.  In particular, we have proven that to obtain the triplet model as a coset, we must include $\AffOthMod{0}$ and $\AffOthMod{1}$ (and hence their twisted images under spectral flow).  These modules do not seem to appear in other discussions relating $\AKMA{sl}{2}_{-1/2}$ and the triplet model (or symplectic fermions).  Second, the level $k=-1/2$ is the simplest example of a variety of fractional level theories whose properties remain largely unexplored.  We do not expect that many other admissible levels have easily guessed free field realisations, so it makes sense to develop methods and techniques which rely on such realisations as little as possible.  Finally, fusing $\AffOthMod{0}$ and $\AffOthMod{1}$ should lead to indecomposable $\AKMA{sl}{2}$-modules whose algebraic structures can be explored in detail.  These indecomposables should provide analogues of the well-known staggered modules of the Virasoro algebra \cite{RohRed96,RidSta09}.  Elucidation of their general properties will be essential in unravelling further properties of fractional level theories.

\subsection{Modular Invariance}

We turn now to a brief discussion concerning the modular invariance of the augmented theory $\AKMA{sl}{2}_{-1/2}^{\text{aug.}}$.  The spectrum of our original $\AKMA{sl}{2}_{-1/2}$-theory was already known to be modular invariant with a perfectly satisfactory Verlinde formula (\secref{secOld}).  We should therefore explain why augmenting the theory by $\AffOthMod{0}$ and $\AffOthMod{1}$ does not destroy this rather nice state of affairs.  At first glance, the situation appears hopeless.  The characters \eqref{eqnCharEType} of $\AffOthMod{0}$ and $\AffOthMod{1}$ do not converge for any value of $z$, so making any sense of them beyond formal generating functions for weight multiplicities seems pointless.  However, there is one possibility which accords well with the mathematical principles described in \secref{secOld} and which also preserves the modular structure found for $\AKMA{sl}{2}_{-1/2}$.

It is perhaps best to first describe this possibility for one of the indecomposable modules studied in \cite{LesLog04}, $\AffOthMod{1/2}^-$ say.  The corresponding character is easy to compute and is in fact also given by \eqnref{eqnCharEType} (with $\lambda = \tfrac{1}{2}$).  It too does not converge for any value of $z$.  However, as remarked above, $\AffOthMod{1/2}^-$ has a submodule generated by the zero-grade state of weight $\tfrac{1}{2}$ and it is not hard to see that this submodule must be isomorphic to $\AffIrrMod{-1/2}^* = \tfunc{\gamma^{-1}}{\AffIrrMod{0}}$.  Moreover, the quotient by this submodule can only be $\AffIrrMod{-3/2} = \tfunc{\gamma}{\AffIrrMod{1}}$.  It therefore follows that, at the level of formal (normalised) characters,
\begin{equation}
\nch{\AffOthMod{1/2}^-}{y ; z ; q} = \nch{\tfunc{\gamma^{-1}}{\AffIrrMod{0}}}{y ; z ; q} + \nch{\tfunc{\gamma}{\AffIrrMod{1}}}{y ; z ; q}.
\end{equation}
As modular characters, where we forget about convergence, the left hand side makes no sense.  But the right hand side does, as the sum of the modular characters of $\tfunc{\gamma^{-1}}{\AffIrrMod{0}}$ and $\tfunc{\gamma}{\AffIrrMod{1}}$.\footnote{Of course, the intersection of the annuli of convergence \eqref{eqnConvergenceRegion} of the corresponding characters is \emph{empty}.  But that is the point of modular characters --- to forget completely about regions of convergence.}  It is therefore natural to define the left hand side to be this sum also.  And of course, this sum gives \emph{zero} because of \eqnref{eqnGenKer} (and the surrounding discussion).

The above prelude is intended to convince the reader that it is natural to assign the modular character $0$ to the module $\AffOthMod{1/2}^-$.  We now want to extend this argument to the irreducible modules $\AffOthMod{0}$ and $\AffOthMod{1}$ which cannot be decomposed into a non-trivial submodule and its quotient.  Mathematically, the above argument amounts to splitting the (otherwise horribly divergent) sum
\begin{equation}
\sum_{n \in \ZZ + \lambda / 2} z^{2n}
\end{equation}
into two pieces, one with $n \geqslant n_0$ and one with $n \leqslant n_0 - 1$.  The cutoff $n_0$ in fact depended upon the grade in the module (power of $q$ in the character).  And of course if one completely ignores the fact that the two sums thereby obtained converge on disjoint sets, then summing gives $0$, independent of the chosen cutoff.

The proposal is then that the irreducible modules $\AffOthMod{0}$ and $\AffOthMod{1}$ should also be assigned modular character $0$.  A consequence is that these modules join those of \eqnref{eqnGenKer} in generating the kernel of the projection from the fusion ring onto the Grothendieck ring of modular characters, thereby preserving the modular structure of \secref{secOld}.  This might seem unpalatable to some, but we believe that this is the only logical way of extending the notion of modular invariance to our augmented theory.  Of course, that this makes a modicum of sense does not guarantee that this proposal has any physical relevance, but at least we can be content in knowing that it preserves the mathematical structures, in particular the Verlinde formula, that have contributed so much to the recent interplay between mathematics and physics.

\subsection{Future Work}

We conclude with a brief outlook.  The most immediate questions raised by the work reported here involve ``fleshing out'' our proposed augmentation to the $\AKMA{sl}{2}_{-1/2}$-theory.  This means computing the fusion rules of the augmented theory and work is already underway in this direction \cite{RidFus10}.  As noted already, we expect that the results of these fusion computations will involve indecomposable $\AKMA{sl}{2}$-modules on which $L_0$ acts non-diagonalisably.  The investigation of the mathematical structure of these indecomposables is therefore also of immediate interest and we will also report on this soon.  We further remark that fusion computations such as these should be of significant interest to those who study string theory on $\func{\group{SL}}{2 ; \RR}$ and its universal cover $\group{AdS}_3$.

It is perhaps worthwhile emphasising that a consequence of the work reported here is a new construction of the triplet algebra.  This is a well known example of what are generally referred to as \emph{W-algebras} in the \cft{} literature --- chiral algebras generated by fields with conformal dimension greater than $2$.  Such algebras are not universal enveloping algebras of Lie algebras, so we know comparatively little about their structure and representations, except in a few isolated cases (see \cite{BouW3A96} for an example).  Various W-algebras have recently been proposed (but without specifying the algebraic structure!) as extended chiral algebras for many \lcfts{} \cite{FeiLog06,RasWEx08,GabFus09}.  It is therefore evident that a construction of such algebras from a much simpler algebra, analogous to obtaining the triplet algebra from $\AKMA{sl}{2}$, would be extremely desirable.  We hope that the ideas outlined here will be of some use in obtaining such constructions.

Another pressing matter is to lift our conclusions from chiral considerations to the bulk regime.  To our knowledge, this question has not yet been seriously addressed for admissible level theories such as $\AKMA{sl}{2}_{-1/2}^{\text{aug.}}$.  Even for the original $\AKMA{sl}{2}_{-1/2}$-theory of \secref{secOld}, the coupling of holomorphic and antiholomorphic sectors is obscured by the fact that the modular invariant only describes a small quotient of the fusion ring.  We expect that the work of \cite{GabLoc99,GabFro08} on such couplings for the triplet model will be very useful aids in this regard.  Finally, we mention that it might be very interesting to consider more stringent constraints upon the consistency of a proposed \cft{}, such as crossing symmetry (or more general sewing constraints).  In particular, it is important to confirm that the two $\AKMA{sl}{2}_{-1/2}$-theories discussed here are indeed consistent \cfts{} (or not, as the case may be!).

Finally, it has recently been brought to our attention that a $\AKMA{u}{1}$ coset of the $\AKMA{sl}{2}_{-4/3}$ theory of \cite{GabFus01} has been studied in the vertex algebra literature \cite{AdaCon05}.  The level $-\tfrac{4}{3}$ theory has $c = -6$, so the coset theory shares its central charge $c = -7$ with another triplet theory, the so-called $\left(1,3\right)$ triplet model (the original triplet model of \cite{GabInd96} corresponds to the $\left(1,2\right)$ theory in this framework).  Nevertheless, it is predicted in \cite{AdaCon05} that the chiral algebra of the coset theory $\AKMA{sl}{2}_{-4/3} / \AKMA{u}{1}$ is generated by a single field of conformal dimension $5$.  The $\left(1,3\right)$ triplet algebra has \emph{three} generators of this dimension, so this prediction is at odds with what one might expect from the results presented here.  It would be extremely interesting to understand why the $k = -\tfrac{4}{3}$ case is different from the $k = -\tfrac{1}{2}$ theory in this regard.

Whatever the outcomes, it should now be clear that the world of admissible, fractional level \WZW{}-theories is ready to be explored.  The technology used here and in \cite{GabFus01,RidSL208,RidFus10} will greatly improve our knowledge of these important models.  We therefore envisage their further study, emphasising that these models should provide basic building blocks for quasi-rational and \lcfts{}, just as the integer-level \WZW{}-models do for rational \cfts{}.

\section*{Acknowledgements}

I would like to thank Pierre Mathieu and Yvan Saint-Aubin for discussions and detailed comments on various drafts of this paper.  I likewise thank Hubert Saleur for his comments and Thomas Creutzig, Matthias Gaberdiel, Thomas Quella and Alexei Semikhatov for useful and enjoyable conversations related to the work presented here.  I also thank Antun Milas for bringing various results from the vertex algebra literature to my attention.

\appendix

\section{A Combinatorial Identity} \label{appComb}

In this appendix, we prove \eqnref{eqnCombId1} for $n \geqslant 0$.  The result for $n < 0$ then follows from symmetry under $n \leftrightarrow -n$.  This proof will follow from an identity of Cauchy \cite[Thm.~2.1]{AndThe76}:
\begin{equation} \label{eqnCauchy1}
\frac{\qnum{az}{\infty}}{\qnum{z}{\infty}} = \sum_{j=0}^{\infty} \frac{\qnum{a}{j}}{\qnum{q}{j}} z^j.
\end{equation}
Here, $\qnum{a}{j}$ denotes the usual $q$-factorial $\prod_{i=0}^{j-1} \brac{1-aq^i}$.  Setting $a=0$ and $z=q^k$ then gives
\begin{equation} \label{eqnCauchy2}
\frac{1}{\qnum{q^k}{\infty}} = \sum_{j=0}^{\infty} \frac{q^{jk}}{\qnum{q}{j}}.
\end{equation}

As is often the case, it is easier to prove a generalisation of \eqnref{eqnCombId1}.  We therefore consider
\begin{align}
\sum_{m=0}^{\infty} \frac{q^m}{\qnum{q}{m}} \frac{\qnum{a}{m+2n}}{\qnum{q}{m+2n}} &= \frac{\qnum{a}{\infty}}{\qnum{q}{\infty}} \sum_{m=0}^{\infty} \frac{q^m}{\qnum{q}{m}} \frac{\qnum{q^{m+2n+1}}{\infty}}{\qnum{aq^{m+2n}}{\infty}} & & \notag \\
&= \frac{\qnum{a}{\infty}}{\qnum{q}{\infty}} \sum_{m=0}^{\infty} \frac{q^m}{\qnum{q}{m}} \sum_{j=0}^{\infty} \frac{\qnum{q/a}{j}}{\qnum{q}{j}} a^j q^{j \brac{m+2n}} & &\text{by (\ref{eqnCauchy1})} \notag \\
&= \frac{\qnum{a}{\infty}}{\qnum{q}{\infty}} \sum_{j=0}^{\infty} \frac{\qnum{q/a}{j}}{\qnum{q}{j}} a^j q^{2 j n} \sum_{m=0}^{\infty} \frac{q^{m \brac{j+1}}}{\qnum{q}{m}} & & \notag \\
&= \frac{\qnum{a}{\infty}}{\qnum{q}{\infty}} \sum_{j=0}^{\infty} \frac{\qnum{q/a}{j}}{\qnum{q}{j} \qnum{q^{j+1}}{\infty}} a^j q^{2 j n} & &\text{by (\ref{eqnCauchy2})} \notag \\
&= \frac{\qnum{a}{\infty}}{\qnum{q}{\infty}^2} \sum_{j=0}^{\infty} q^{2 j n} \prod_{i=1}^j \brac{a-q^i}. & &
\end{align}
Putting $a=0$ then gives
\begin{equation}
\sum_{m=0}^{\infty} \frac{q^m}{\qnum{q}{m} \qnum{q}{m+2n}} = \frac{1}{\qnum{q}{\infty}^2} \sum_{j=0}^{\infty} \brac{-1}^j q^{j \brac{j+1} / 2 + 2jn}.
\end{equation}
Replacing $m$ by $m-n$ on the left hand side and $j$ by $m-2n$ on the right hand side then gives \eqnref{eqnCombId1}.

\section{Non-Highest Weight $\SLA{sl}{2}$-Modules} \label{appSL2}

Let $\ket{v_m}$ be an eigenstate of $h \in \SLA{sl}{2}$ with eigenvalue $m$.  If we assume that $\ket{v_m}$ is not highest weight, then we may write
\begin{equation} \label{eqnDefAlpha}
\ket{v_{m + 2}} = e \ket{v_m} \qquad \text{and} \qquad f \ket{v_m} = \alpha_{m} \ket{v_{m - 2}},
\end{equation}
for some constant $\alpha_{m}$.  Repeating this for $\ket{v_{m + 2}}$, $\ket{v_{m - 2}}$ and the states thereby generated, we construct a basis for the module generated by $\ket{v_m}$.  Note however that $e^{\dag} = f$ gives
\begin{equation} \label{eqnNormRel}
\braket{v_m}{v_m} = \bracket{v_{m - 2}}{f}{v_m} = \alpha_{m} \braket{v_{m - 2}}{v_{m - 2}},
\end{equation}
so $\alpha_{m} \in \RR$ whenever $\braket{v_m}{v_m}$ (and $\braket{v_{m - 2}}{v_{m - 2}}$) are non-zero.  Moreover,
\begin{equation}
m \ket{v_m} = \brac{fe-ef} \ket{v_m} = \brac{\alpha_{m+2} - \alpha_{m}} \ket{v_m},
\end{equation}
which gives
\begin{equation} \label{eqnAlphaRel}
\alpha_{m+2} = \alpha_{m} + m.
\end{equation}
We remark that $m \notin \RR$ would now imply that $\alpha_{m} \notin \RR$, hence that $\braket{v_m}{v_m} = 0$.  This reflects the fact that one cannot define a non-trivial hermitian form on an $\SLA{sl}{2}$-module with non-real weights.

Recall that the quadratic Casimir is defined (up to normalisation) by inverting the Killing form \eqref{eqnKilling}:
\begin{equation}
Q = \tfrac{1}{2} h^2 - ef - fe.
\end{equation}
Using \eqnref{eqnAlphaRel}, we calculate its eigenvalue on $\ket{v_m}$ to be
\begin{equation} \label{eqnQEig}
Q_{m} = \tfrac{1}{2} m^2 - \alpha_{m} - \alpha_{m+2} = \tfrac{1}{2} m^2 - m - 2 \alpha_{m},
\end{equation}
and it is easily checked (using \eqnref{eqnAlphaRel} again) that $Q_{m}$ is periodic in $m$ with period $2$.  This eigenvalue is therefore constant on the $\SLA{sl}{2}$-module generated by $\ket{v_m}$, as it must be.

\end{document}